\documentclass[twocolumn]{aastex62}
\usepackage{enumitem}
\usepackage{graphicx,url}
\usepackage{color}
\usepackage{amsmath}
\usepackage{amssymb}
\usepackage{mathtools}
\usepackage[htt]{hyphenat}
\usepackage[utf8]{inputenc}

\def\j1550{XTE~J1550$-$564}

\def\u1630{4U~1630$-$47}

\def\rxte{{\it RXTE}}

\def\swift{{\it Swift}}
\def\nustar{{\it NuSTAR}}

\accepted{by ApJ, 2021 January 15}

\shorttitle{Reflection Spectroscopy of 4U~1630$-$47}
\shortauthors{Connors et al.}

\setlist[itemize]{leftmargin=*}

\begin{document}

\title{Reflection Modeling of the Black Hole Binary 4U~1630$-$47: the Disk Density and Returning Radiation}

\correspondingauthor{Riley~M.~T.~Connors}
\email{rconnors@caltech.edu}

\author{Riley~M.~T.~Connors}
\affiliation{Cahill Center for Astronomy and Astrophysics, California Institute of Technology, \\
  Pasadena, CA 91125, USA}
  
 \author{Javier~A.~Garc\'ia}
\affiliation{Cahill Center for Astronomy and Astrophysics, California Institute of Technology, \\
  Pasadena, CA 91125, USA}
  \affiliation{Dr Karl Remeis-Observatory and Erlangen Centre for Astroparticle Physics,\\
  Sternwartstr. 7, D-96049 Bamberg, Germany}
  
  \author{John Tomsick}
\affiliation{Space Sciences Laboratory, University of California Berkeley,\\
7 Gauss Way, Berkeley, CA 94720-7450}

\author{Jeremy~Hare}
\affiliation{NASA Goddard Space Flight Center, Greenbelt, MD 20771, USA}

 \author{Thomas Dauser}
\affiliation{Dr Karl Remeis-Observatory and Erlangen Centre for Astroparticle Physics,\\
  Sternwartstr. 7, D-96049 Bamberg, Germany}

\author{Victoria Grinberg}
\affiliation{Institut f\"ur Astronomie und Astrophysik, Universit\"at T\"ubingen, Sand 1, 72076 T\"ubingen, Germany}

\author{James~F.~Steiner}
\affiliation{MIT Kavli Institute, 77 Massachusetts Avenue, 37-241, \\
Cambridge, MA 02139, USA}
\affiliation{CfA, 60 Garden St. Cambridge, MA 02138, USA}

\author{Guglielmo~Mastroserio}
\affiliation{Cahill Center for Astronomy and Astrophysics, California Institute of Technology, \\
  Pasadena, CA 91125, USA}

\author{Navin Sridhar}
\affiliation{Department of Astronomy, Columbia University,\\
    550 W 120th St, New York, NY 10027, USA}

\author{Andrew~C.~Fabian}
\affiliation{Institute of Astronomy, University of Cambridge, Madingley Road, Cambridge CB3 0HA, UK}

\author{Jiachen~Jiang}
\affiliation{Department of Astronomy, Tsinghua University, Shuangqing Road, Beijing 100084, China}

\author{Michael~L.~Parker}
\affiliation{European Space Agency (ESA), European Space Astronomy Centre (ESAC), E-28691 Villanueva de la Ca\~nada, Madrid, Spain}

\author{Fiona Harrison}
 \affiliation{Cahill Center for Astronomy and Astrophysics, California Institute of Technology, \\
  Pasadena, CA 91125, USA}
  
 \author{Timothy~R.~Kallman}
 \affiliation{NASA Goddard Space Flight Center, Greenbelt, MD 20771, USA}

  

\begin{abstract}

We present the analysis of X-ray observations of the black hole binary \u1630\ using relativistic reflection spectroscopy. We use archival data from the \rxte, \swift, and \nustar\ observatories, taken during different outbursts of the source between $1998$ and $2015$. Our modeling includes two relatively new advances in modern reflection codes: high-density disks, and returning thermal disk radiation. Accretion disks around stellar-mass black holes are expected to have densities well above the standard value assumed in traditional reflection models (i.e., $n_{\rm e}\sim10^{15}~{\rm cm^{-3}}$). New high-density reflection models have important implications in the determination of disk truncation (i.e., the disk inner radius). This is because one must retain self-consistency in the irradiating flux and corresponding disk ionization state, which is a function of disk density and system geometry. We find the disk density is $n_{\rm e}\ge10^{20}~{\rm cm^{-3}}$ across all spectral states. This density, combined with our constraints on the ionization state of the material, implies an irradiating flux impinging on the disk that is consistent with the expected theoretical estimates. Returning thermal disk radiation---the fraction of disk photons which bend back to the disk producing additional reflection components---is expected predominantly in the soft state. We show that returning radiation models indeed provide a better fit to the soft state data, reinforcing previous results which show that in the soft state the irradiating continuum may be blackbody emission from the disk itself.

\end{abstract}


%
%
%
\section{Introduction}\label{sec:intro}
Black hole X-ray binaries (BHBs) with low-mass companions are known to show transient behavior, with many ($8\mbox{--}9$) orders of magnitude variation in luminosity during outburst. During such outbursts, BHBs display various states, characterized via several different criteria, most notably the degree of X-ray variability and X-ray spectral properties (see, e.g., \citealt{Homan2005}). 

Generally, one can characterize the typical spectral states as follows: hard state, steep power-law state (a part of the broadly classified intermediate state), and thermal-dominated or soft state (see, e.g., \citealt{rm06}). BHBs brighten out of their quiescent phase into the hard state, typically, exhibiting a power-law-like, hard spectrum with $\Gamma\sim1.5\mbox{--}2$.  The source will then typically transition either directly to the soft, thermal-dominated state, in which a multitemperature blackbody spectrum is typically observed at $kT\sim1~{\rm keV}$, or first through the steep power-law state---which lies within the broadly defined intermediate state. The steep power-law state is characterized by comparably strong disk blackbody and power-law emission with a power-law index of $\Gamma\sim2.5\mbox{--}3$. 

Physical interpretations of the totality of BHB outburst evolution are many, but it is well-established that the general transient nature of BHBs is a result of viscous-thermal instabilities in the accretion flow, known as the disk instability model (DIM; \citealt{Lasota2001}). The sharp increase in luminosity during an outburst is associated with a rapid increase in the accretion rate $\dot{M}$ of a geometrically thin, optically thick disk progressing from the outer to inner flow. The structure of this accretion disk was originally characterized by \cite{SS1973}, emitting strongly as a multitemperature blackbody at high accretion rates. Thus, the intuitive picture of the outburst of a BHB is one of the approach of this thin disk towards the innermost stable circular orbit (ISCO; \citealt{Novikov1973}), which is associated with maximal conversion of gravitational to kinetic/thermal/radiative energy---the thermal-dominated state \citep{McClintock2006}.

 Modeling of the reflected X-ray spectra \citep{Ross2005,Ross2007,Garcia2014,Dauser2014} of BHBs has proven a useful tool in our attempts to understand physical changes to the accretion flow across spectral states (see \citealt{Garcia2015} and references therein). To this end, recent developments to reflection models have allowed us to make progress in reflection studies (\citealt{Garcia2016,Tomsick2018,JJiang2019}, Mastroserio et al., in preparation). As the disk density increases beyond the previously assumed value of $n_{\rm e}=10^{15}~\mathrm{cm^{-3}}$, additional free-free heating occurs in the upper layers of the disk, leading to a strong quasi-thermal component with a temperature that is proportional to both the density and ionization of the disk (and thus the irradiating flux, since the ionization parameters is given by $\xi=4 \pi F_{\rm irr} / n_{\rm e}$, where $F_{\rm irr}$ is the ionizing flux, and $n_{\rm e}$ is the disk density; \citealt{Garcia2016}). High density models have been applied in several recent works to the X-ray spectra of BHBs, showing systematic effects on key physical model parameters \citep{Tomsick2018,JJiang2019}. For example, \cite{JJiang2019} found a weakening of disk truncation constraints when comparing with the very low values found by \citep{Wang-Ji2018} during the hard state of GX~339$-$4, though the differences were not large. There has also been success applying the same high-density reflection models to AGN spectra \citep{JJiang2019b,Garcia2019}. 
 
 There is still work needed to characterize the evolution of BHB reflection properties, and their dependence on the disk density, during the transition from the hard to soft states. Recent work has focused on tracking the key accretion flow properties of BHBs with advanced reflection models (e.g., \citealt{Sridhar2020,Connors2020}), showing that as BHBs transition from the hard to the soft state, the illuminating continuum may be evolving from a power-law like coronal spectrum to a more disk-blackbody-like spectrum due to the luminosity of the disk. Thus, a comprehensive picture of the evolving accretion flow during transition and into the soft state requires a full treatment of both density effects and the impact of returning disk emission. 
 
 \u1630 is one of the most active BHBs to have been observed. Discovered first by the {\it Vela 5B} satellite in 1969 \citep{Priedhorsky1986}, it has since been detected in outburst $>20 $ times \citep{Tetarenko2016}. It has a high hydrogen column density along the line of sight, with $N_{\rm H}\sim(4\mbox{--}12)\times10^{22}~{\rm cm^{-2}}$ \citep{Kuulkers1998,Tomsick1998}, with more recent high-resolution X-ray grating spectroscopy with {\it Chandra} indicating a likely value close to $10^{23}~{\rm cm^{-2}}$ \citep{Gatuzz2019}. Though the dynamical mass of the BH has not been determined, an indirect mass estimate of $\sim10~M_{\odot}$ was obtained via scalings of the photon index, $\Gamma$, with accretion rate \citep{Seifina2014}. IR observations of its 1998 outburst led to a distance estimate of $D\sim10\mbox{--}11~{\rm kpc}$ \citep{Augusteijn2001,Seifina2014}, and a rough estimate for the orbital period of a few days with an early-type companion star, which is comparatively high. Dips observed in the Proportional Counter Array (PCA; \citealt{Jahoda1996}) data from the {\it Rossi X-ray Timing Explorer} (\rxte) showed that the source must have a high binary inclination ($i>60^{\circ}$; \citealt{Tomsick1998}), and later scaling estimates suggest the inclination is probably $<70^{\circ}$ \citep{Seifina2014}. Indeed recent reflection modeling showed a preference for an inclination of $\sim64^{\circ}$ \citep{King2014}, though this comes with the caveat that reflection spectra depend on the disk inclination, which could in principle be misaligned with the orbital inclination. 
 
 \u1630 also exhibits a strong ionized disk wind in the soft state \citep{Kubota2007,Hori2014,King2014,DiazTrigo2014,Hori2018,Gatuzz2019}, the primary signatures of which are two prominent absorption lines from \ion{Fe}{25} and \ion{Fe}{26} with moderate shifts observed by high-resolution X-ray spectroscopy (provided by observations made by {\it Chandra}, {\it XMM-Newton} and {\it Suzaku}). 
 
 In this paper we focus on several different observations of \u1630, and perform detailed reflection modeling of spectra from the (hard)intermediate to soft states in order to characterize the inner accretion flow during the transition. The key goals are to understand the systematic differences between reflection from low-density and high-density disks. Our focus is primarily on the interplay between the intrinsic disk blackbody and coronal emission, and the reprocessed irradiating flux, a component which depends strongly on density. We use the spectral evolution to constrain this interplay, given the shift in the dominance of the coronal and disk components during the transition.  
 
 
 The structure of this paper as as follows: in Section~\ref{sec:data} we present all the observations we select for modeling. These include \rxte-PCA, the {\it Nuclear Spectrscopic Telescope Array} (\nustar; \citealt{Harrison:2013}), and {\it Neil Gehrels Swift Observatory} (\swift; \citealt{Gehrels2004}). In Section~\ref{sec:modeling} we describe our modeling setup and present results of reflection modeling to PCA and \nustar/\swift-XRT spectra. In Section~\ref{sec:discussion} we discuss the implications of our reflection fitting results and give our conclusions.

\section{Data} \label{sec:data}

We select observations of \u1630 made by \rxte, \nustar, and \swift, specifically its X-ray Telescope (XRT; \citealt{Krimm2013}). In the following subsections we outline the details of those observations and how we reduced the data. 

\subsection{\rxte} \label{subsec:rxte}
\begin{figure}
\centering
\includegraphics[width=\linewidth]{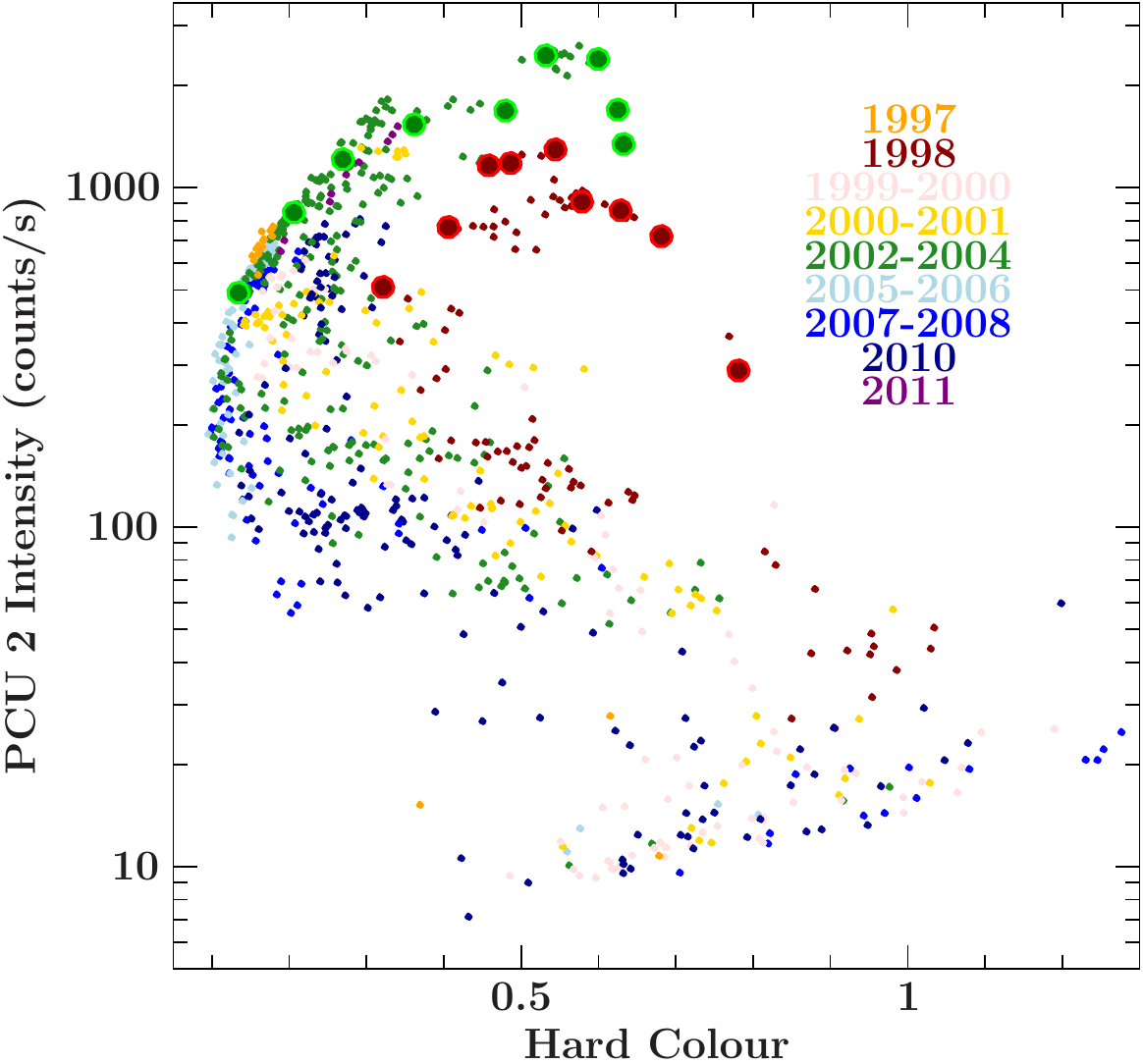}
\includegraphics[width=\linewidth]{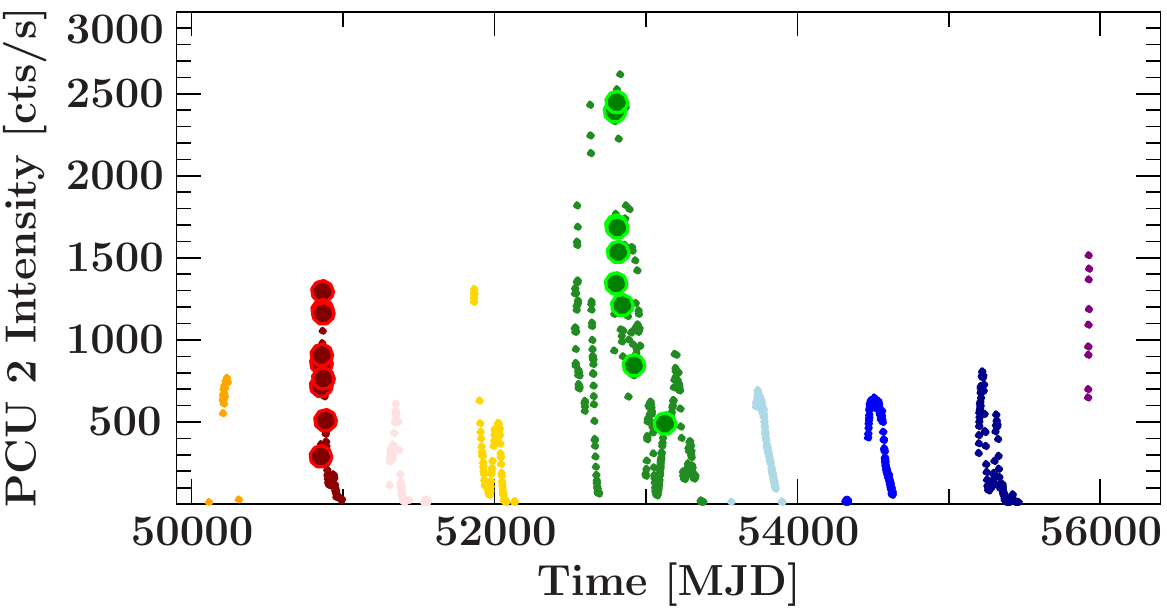}
\caption{\rxte-PCA hardness-intensity diagram (top) and light curve (bottom), including all \rxte\ observations over roughly 10 years---9 outbursts in total. Hard color is defined as the ratio of source counts in the $8.6\mbox{--}18$ keV (hard) and $5\mbox{--}8.6$~keV (soft) energy bands. The selected observations of outbursts 2 and 5, as shown in Table~\ref{tab:rxte}, are highlighted as larger red and green points.}
\label{fig:rxte_data}
\end{figure}

\rxte\ observed \u1630 during a total of 9 outbursts between the years of 1996 and 2012, when \rxte\ ceased operating. The short recurrence time and broad coverage with \rxte\ make this a rich dataset, with $>1000$ PCA spectra. The full lightcurve and hardness-intensity diagram (HID) are shown in Figure~\ref{fig:rxte_data}. The HID shows that \u1630 spends the majority of its outbursts in either the soft or intermediate state (including brighter intermediate states akin to the steep power law state, or very high state), which is likely a combination of rapid rise times during outburst, and possibly a skew to softer emission due to a hotter, optically thick disk. However, \rxte\ caught \u1630 in hard and hard-intermediate states during outbursts in $1998$ and $2002\mbox{--}2004$ (an unusually long and multi-peaked outburst, see \citealt{Tomsick2005}). 

We select data from these two outbursts (see Table~\ref{tab:rxte}, and highlighted data in Figure~\ref{fig:rxte_data}) in order to characterize the hard-to-soft transitions of \u1630, and make direct comparisons with the more recent \nustar\ and \swift\ observations. The selection criteria is such that we cover a range of hardness whilst maximizing the total number of detector counts. The key goals of modeling both these datasets are: to constrain the disk density as the source is transitioning; build a consistent picture of the structure of the inner accretion flow during transition; and investigate the key effects the assumption of a higher disk density has on other key disk, coronal, and reflection model parameters. 

\begin{deluxetable}{lccccr}
\tabletypesize{\scriptsize}
\tablecaption{Properties of the selected \rxte-PCA (PCU~2) X-ray spectra from the $1998$ and $2002\mbox{--}2004$ outbursts of \u1630.}
\tablecolumns{6}
\tablehead{
\colhead{ObsID} & 
\colhead{MJD} & 
\colhead{HR\tablenotemark{a}} & 
\colhead{State\tablenotemark{b}} &
\colhead{$N_{\rm counts}$\tablenotemark{c}} & 
\colhead{cts~${\rm s^{-1}}$ \tablenotemark{d}} \\
& & & & ($10^6$) 
}
\startdata
\hline
& & & 1998 & & \\
\hline
30178-01-01-00 & 50853.1 & 0.78 & Hard & $0.3$ & 289 \\
30178-02-01-00 & 50855.0 & 0.67 & INT & $0.9$ & 719 \\
30178-02-01-01 & 50856.9 & 0.63 & INT & $2.6$ & 857 \\
30178-02-02-01 & 50858.8 & 0.58 & INT & $7.9$ & 909 \\
30178-02-13-00 & 50864.2 & 0.54 & SPL & $1.7$ & 1295 \\
30178-01-12-00 & 50864.6 & 0.49 & SPL & $1.9$ & 1181 \\
30188-02-18-00 & 50868.6 & 0.46 & SPL & $4.4$ & 1162 \\
30188-02-19-00 & 50869.7 & 0.41 & Soft & $1.3$ & 765 \\
30172-01-01-00 & 50883.8 & 0.32 & Soft & $3.5$ & 509 \\
\hline
& & & $2002\mbox{--}2004$ & & \\
\hline
80117-01-03-00G & 52795.3 & 0.60 & SPL & $18$ & 2392 \\
80117-01-06-00 & 52802.8 & 0.63 & INT & $11$ & 1344 \\
80117-01-07-00 & 52804.2 & 0.62 & INT & $4$ & 1696 \\
80117-01-07-01 & 52806.5 & 0.53 & SPL & $6$ & 2450 \\
80117-01-08-00 & 52810.1 & 0.48 & SPL & $2.2$ & 1685 \\
80117-01-09-01 & 52816.3 & 0.36 & Soft & $2.7$ & 1536 \\
80117-01-13-02 & 52843.4 & 0.27 & Soft & $2.4$ & 1212 \\
80420-01-07-01 & 52920.9 & 0.21 & Soft & $1.0$ & 845 \\
90410-01-03-00 & 53125.5 & 0.13 & Soft & $1.1$ & 490 \\
\enddata
\tablenotetext{a}{Hardness ratio given by source counts in [$8.6\mbox{--}18$~keV]/[$5\mbox{--}8.6$~keV] bands.}
\tablenotetext{b}{INT$=$intermediate, SPL$=$Steep power law.}
\tablenotetext{c}{Number of counts in the $3\mbox{--}45$~keV band of the PCU~2 spectra.}
\tablenotetext{d}{Total $3\mbox{--}45$~keV count rate.}
\label{tab:rxte}
\end{deluxetable}

\begin{deluxetable}{lccccr}
\tablecaption{Properties of the selected \nustar\ and \swift-XRT spectra.}
\tablecolumns{5}
\tablehead{
\colhead{ObsID} & 
\colhead{Instr.} &
\colhead{MJD} & 
\colhead{Exp.} & 
\colhead{$N_{\rm counts}$\tablenotemark{a}} & 
\colhead{cts~${\rm s^{-1}}$ \tablenotemark{b}} \\
& & & [ks] & ($10^6$) 
}
\startdata
40014009001 & \nustar\ & 56344  & $14.7$ & $1.6$ & 112 \\
90002004002 & \nustar\  & 57073  & $15.7$ & $2.8$ & 176 \\
90002004004 & \nustar\  & 57077  & $16.1$ & $1.8$ & 113 \\
00080510001 & \swift-XRT & 56344  & $2.1$ & $0.01$ & 6 \\
00081434001 & \swift-XRT & 57077  & $1.6$ & $0.02$ & 14 \\
\enddata
\tablenotetext{a}{Number of counts in the $3\mbox{--}79$~keV band of the \nustar-FPMA detector, and $0.5\mbox{--}10$~keV band of the \swift-XRT detector, post-grouping.}
\tablenotetext{b}{Mean $3\mbox{--}79$~keV (\nustar) and $0.5\mbox{--}10$~keV (\swift-XRT) count rates, post-grouping.}
\label{tab:nustar-swift}
\end{deluxetable}

All the data shown in Table~\ref{tab:rxte} are publicly available on the \rxte\ archive via HEASARC (High Energy Astrophysics Science Archive Research Center). We extracted the PCA data only, removing data lying within $10$~min of the South Atlantic Anomaly (SAA). We use data from proportional counter unit (PCU) 2, due to its superior calibration and extensive coverage (all PCA exposures include PCU~2 data). We corrected the PCU~2 data using the publicly available tool {\tt pcacorr} \citep{Garcia2014b}, and introduce $0.1$\% systematic errors to all channels. The level of systematic errors to impose on the data is based upon the direction of \cite{Garcia2014b}, showing that the reduction in systematics achieved by the {\tt pcacorr} tool allows one to lower the assumed systematics to $\sim0.1$\%. We further group the PCU~2 spectra at a signal-to-noise ratio of 4, which achieves sufficient oversampling of the source counts to outweigh the background at high energies. We ignore PCU~2 data in channels $1\mbox{--}4$, and beyond $45$~keV.

\subsection{\nustar\ and \swift} \label{sec:nustar_data}


We use three archival \nustar\ observations of 4U 1630-47 taken while the source was in the soft state. {\sl NuSTAR} first observed 4U 1630-47 on UT 2013 February 21 with a total exposure time of 14.7 ks (ObsID 40014009001). 
The two additional {\sl NuSTAR} observations of 4U 1630-47 taken on 2015 February 20 and 24 (ObsIDS 90002004002 and 90002004004, respectively) have exposure times of 15.7 ks and 16.1 ks respectively. 
All three {\sl NuSTAR} observations were reduced using version 1.8.0 of the {\sl NuSTAR} Data Analysis Software (NuSTARDAS) package and version 20181030 of the calibration database (CALDB). The {\sl NuSTAR} source spectra for both focal plane modules (FPMA/B) were extracted, using the standard procedures, from an $r=80''$ circular region centered on the source. The background spectra were extracted from a source-free circular region ($r\approx70''$) placed on a separate detector chip and avoiding the stray light visible in all three observations. After reducing the data there is some moderate variability remaining (particularly in the first two observations), which may introduce some noisiness in the spectral fits, however, the rms variability on the count rate is on the order of just $10\%$ during each exposure, thus we use the full time-averaged spectra to maximize our statistics. The spectra from all three {\sl NuSTAR} observations were grouped to have a signal-to-noise ratio of at least 10 per energy bin prior to fitting, and we include data only in the $3\mbox{--}79$~keV band. The properties of the data are summarized in Table~\ref{tab:nustar-swift}.

The {\it Neil Gehrels {\sl Swift}}-XRT \citep{Burrows2005} simultaneously observed 4U 1630-47 with {\sl NuSTAR} on two occasions. The first simultaneous {\sl Swift}-XRT observation (ObsID 00080510001), lasting 2 ks and taken in photon counting (PC) mode, was coincident with {\sl NuSTAR} ObsID 40014009001. The source spectrum was extracted from an inner annulus between $30''\mbox{--}90''$, while the background spectrum was extracted from an outer annulus between $200''\mbox{--}300''$.  The second simultaneous {\sl Swift}-XRT observation (ObsID 00081434001), lasting 1.6 ks and taken in window timing (WT) mode, was coincident with {\sl NuSTAR} ObsID 90002004004. The source spectrum was extracted from an annulus centered on the source having an inner radius of 4 pixels and outer radius of 20 pixels. The inner pixels were excluded from the spectral extraction region to eliminate the effects of pile-up in the PSF core (see e.g., \citealt{Romano2006}). The background spectrum was extracted from an annulus centered on the source with inner and outer radii of 90 and 110 pixels. {\sl Swift} observation 00081434001 was reduced using HEASOFT version 6.25, observation 00080510001 with HEASOFT version 6.26.1, and both with the x20190412 version of the CALDB. Prior to fitting, the {\sl Swift} spectra were grouped to have a signal-to-noise ratio of at least 10 (00080510001) and 15 (00081434001) per energy bin, and we include data only in the $0.5\mbox{--}10$~keV band. The properties of the data are summarized in Table~\ref{tab:nustar-swift}.



\section{Modeling} \label{sec:modeling}

Our modeling approach to the multiple \rxte-PCA and simultaneous \nustar/\swift\ data of \u1630 is as follows. We fit the data with three classes of models, each consisting of a Comptonized multitemperature disk blackbody spectrum plus reflected emission. The distinction is between the nature of the reflected component, and each can be characterized as follows:

\begin{itemize}
\item Reflection from a low density disk (fixed to $n_{\rm e}=10^{15}~{\rm cm^{-3}}$), due to irradiating coronal IC photons.
\item Reflection from a disk with variable density (up to $n_{\rm e}=10^{22}~{\rm cm^{-3}}$), due to irradiating coronal IC photons.
\item Reflection of returning disk radiation
\end{itemize}

The first two models listed are motivated by our goal to test the effects of the disk density on the resultant physical, steady-state characteristics of the accretion flows of BHBs. The dichotomy between the reprocessed quasi-thermal emission of the irradiated upper layers of BHB accretion disks (captured in reflection models which extend to high densities), and the intrinsic disk blackbody emission, has been highlighted in recent works \citep{Zdziarski2020a,Zdziarski2020b}. 




 One can make a direct comparison of the disk ionization, $\xi$, resulting from fitting reflection models, with the ionization deduced from a rough calculation of the irradiating flux, such that $\xi=4\pi F_{\rm irr}/n_{\rm e}$. We perform these consistency checks with the derived model parameters of our high-density reflection fits. The details of the reflection fits and results are shown in the Section~\ref{sec:pca_fits}. All spectral analysis was performed using Xspec, v12.10.1s \citep{Arnaud1996}.  
 
We then investigate the presence of returning radiation in the soft-state observations of \nustar/\swift-XRT by way of a comparison with high-density reflection fits, showing some of the key differences in parameter constraints (Section~\ref{sec:nustar_fits}). 

\subsection{\rxte-PCA modeling} \label{sec:pca_fits}

As outlined in Section~\ref{subsec:rxte}, we selected a subset of PCA observations from outbursts 2 and 5 of \u1630 (see Figure~\ref{fig:rxte_data}), which occurred in $1998$ and $2002\mbox{--}2004$. We model these selected spectra with both a low-density and high-density reflection component, as well as a Comptonized disk component. 

\begin{deluxetable*}{ll}
\tablecaption{A summary of all models used in spectral fits.} 
\tablecolumns{2}
\tablehead{
\colhead{Model} & 
\colhead{Components} 
}
\startdata
(a) & {\tt TBabs(simplcut$\otimes$diskbb+relconvlp$\otimes$reflionx)} \\
(b) & {\tt TBabs(simplcut$\otimes$diskbb+relconvlp$\otimes$reflionxHD)} \\
(b.2) & {\tt crabcorr*xscat*TBabs(simplcut$\otimes$diskbb + relconv$\otimes$reflionxHD)*xstar} \\
(c) & {\tt crabcorr*xscat*TBabs(simplcut$\otimes$diskbb + relxillNS)*xstar}
\enddata
\tablecomments{{\tt reflionx} and {\tt reflionxHD} differ only in that the latter has a variable disk density, with $n_{\rm e} \le 10^{22}~{\rm cm^{-3}}$.}.
 \label{tab:models}
\end{deluxetable*}

\subsubsection{Model Setup} \label{subsec:model_setup}
The models, (a) and (b), are defined explicitly in Table~\ref{tab:models}. Model (a) represents the standard low-density reflection model, and model (b) is the high-density reflection model. The component {\tt TBabs} accounts for interstellar absorption, implementing the elemental abundance tables of \cite{Wilms2000}. We adopt the \cite{Verner1996} atomic cross sections. The {\tt simplcut} component is a model for inverse-Compton (IC) scattering in a coronal plasma \citep{Steiner2017}, and is an extension of the model {\tt simpl} \citep{Steiner2009}. {\tt simplcut} acts as a convolution model, scattering the disk photons (as described by the multitemperature disk blackbody component {\tt diskbb}; \citealt{Mitsuda1984}) that traverse hot coronal plasma. It uses the IC spectral calculations given by the Xspec model {\tt nthComp} \citep{Zdziarski1996,Zycki1999}. The temperature of the photon pool for scattering is tied to the {\tt diskbb} value. The parameter $f_{\rm sc}$ sets the proportion of photons which are scattered in the corona. Parameterizing the fraction of photons scattered into the power-law component in this way simplifies the two-part problem: in reality some fraction of photons enter the coronal plasma (a covering fraction), and then some portion of those photons undergo scattering (the optical depth, $\tau$, of the corona sets this quantity). The $f_{\rm sc}$ parameter combines these two quantities, by assuming all photons enter the corona. Thus $f_{\rm sc}$ is a lower bound on the optical depth of the corona, where $\tau \ge -\ln(1-f_{\rm sc})$. From a modeling perspective, implementing {\tt simplcut} as a model for IC scattering in a corona is preferable to treating the disk and corona components independently (with their own normalizations) because the former enforces self-consistency in the fitting procedure (the coronal IC component can only exist with some total number of disk photons, thus placing limits on the disk flux in the modeling procedure). 

Models (a) and (b) are identical in regard to the disk and coronal spectral components, but they differ in the nature of the reflection spectrum. Model (a) assumes a disk density fixed at $10^{15}~{\rm cm^{-3}}$. The reflection spectrum is calculated assuming illumination atop the disk by an IC coronal spectrum, as given by the {\tt nthComp} model \citep{Zdziarski1996,Zycki1999}, and thus coincides with the illumination spectrum given by {\tt simplcut$\otimes$diskbb}. 

The model {\tt reflionxHD} is the latest version of the long-standing reflection model {\tt reflionx} \citep{Ross2005,Ross2007}, which operates as a grid of reflection spectra which can be implemented in Xspec to perform statistical modeling of X-ray spectra. {\tt reflionxHD} differs from the original {\tt reflionx} model in two key ways: (i) the irradiating continuum ionizing the disk is based on the {\tt nthComp} Comptonization model, as opposed to the approximation of a cutoff power law; (ii) the disk number density is a model parameter extending from $10^{15}~{\rm cm^{-3}}$ up to $10^{22}~{\rm cm^{-3}}$. The model component {\tt relconvlp} is part of the {\tt relxill} distribution of models \citep{Garcia2014,Dauser2014} for relativistic reflection, and acts as a convolution model to relativistically smear any input spectrum, assuming a lamppost geometry for the irradiating source. Thus the convolution procedure {\tt relconvlp$\otimes$reflionxHD} applies relativistic smearing to the non-relativistic {\tt reflionxHD} model. We adopt {\tt relconvlp}, the lamppost version of the coronal geometry, based upon previous constraints on the dimensionless BH spin, $a_{\star}=0.985$, and a steep emissivity, $q>9$ \citep{King2014}.

\subsubsection{Parameters} \label{subsec:parameters}
In all fits to the PCA data shown in Figure~\ref{fig:rxte_data} and Table~\ref{tab:rxte} we treat the model parameters as follows. Based on explorative preliminary fits to all the data, we fix the hydrogen column density ($N_{\rm H}$, a parameter of the {\tt TBabs} spectral component) to $1.4\times10^{23}~{\rm cm^{-2}}$. This decision was based on small yet impactful differences between the constrained value of $N_{\rm H}$ as a function of spectral state and data counts, and an approximate range of values between $1.3$ and $1.5\times10^{23}~{\rm cm^{-2}}$. We further justify this choice due to the uncertainties on $N_{\rm H}$ in the literature, as measured in the X-ray band \citep{Tomsick1998,Tomsick2000,Gatuzz2019}, as well as the presence of dust scattering along the line of sight \citep{Kalemci2018}. The {\tt simplcut} model has three variable parameters: $\Gamma$, the power law photon index, $f_{\rm sc}$, the scattering fraction, and $kT_{\rm e}$, the coronal thermal electron temperature. The {\tt diskbb} component is characterized by the disk normalization $N_{\rm disk}$, and the inner disk temperature $T_{\rm in}$. The {\tt reflionxHD} model has parameters for the inner disk temperature $T_{\rm in}$ (this sets the input spectrum for IC scattering, see Jiang et al. 2020, submitted to ApJ), photon index $\Gamma$, and electron temperature $kT_{\rm e}$---these are all tied to their respective values as given by {\tt simplcut} and {\tt diskbb}. Further {\tt reflionxHD} parameters are disk ionization $\xi$, iron abundance $A_{\rm Fe}$, density $n_{\rm e}$, and normalization $N_{\rm refl}$. We fix $A_{\rm Fe}=5$ in our fits, again based upon preliminary fits in which typical constrained values were found to be close to $5$, with fairly large uncertainties. 

\begin{figure*}
\centering
\includegraphics[width=0.48\linewidth]{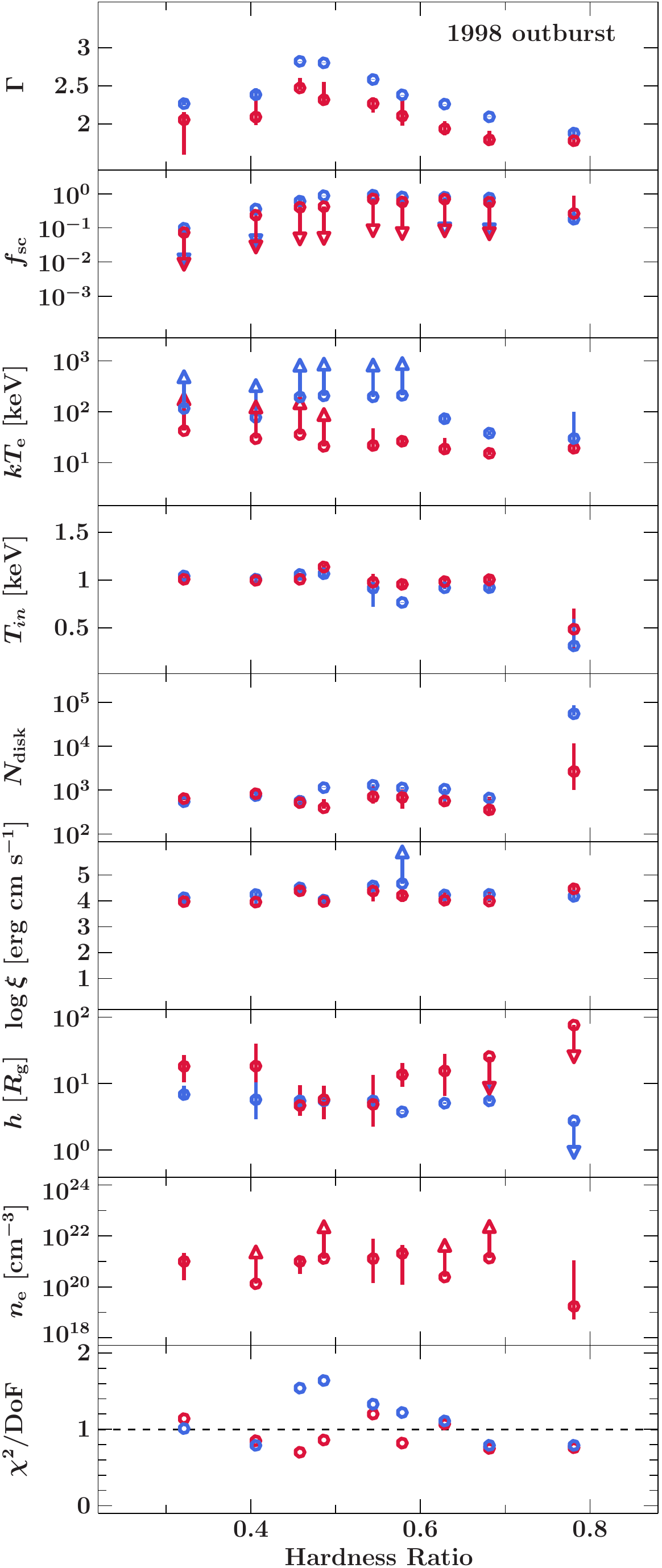}
\includegraphics[width=0.4111\linewidth]{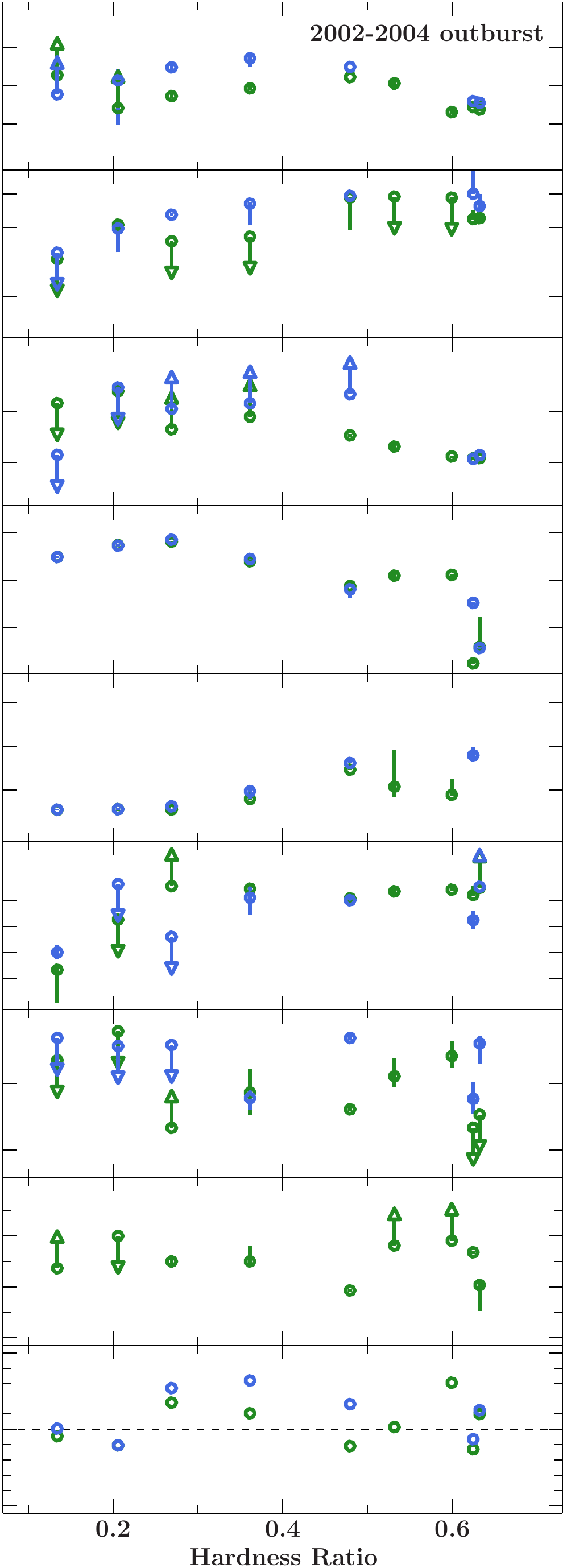}
\caption{Parameter constraints as a function of hardness ratio resulting from modeling the selected \rxte-PCA data from observations of \u1630 during outbursts in $1998$ (left) and $2002\mbox{--}2004$ (right). In both plots the blue points show fits of model (a) with $n_{\rm e}=10^{15}~{\rm cm^{-3}}$. Red and green points show the high-density reflection (Model b) modeling results with data from outbursts in $1998$ and $2002\mbox{--}2004$ respectively. Observations 80117-01-03-00G and 80117-01-07-01 (hardness ratio 0.60 and 0.53) have no low-density (blue) modeling results included in the right hand panel due to very poor fits to those data. }
\label{fig:par_trends}
\end{figure*}

The model {\tt relconvlp} has parameters which pertain to the source geometry: the black hole spin $a_{\star}$, inner disk radius $R_{\rm in}$, outer disk radius $R_{\rm out}$, disk inclination $i$, the lamppost height $h$, and $\Gamma$, the powerlaw index of the IC spectrum. We fix the black hole spin to maximal, $a=0.998$, and inclination to $i=64^{\circ}$, based on the reflection modeling results of \cite{King2014}, which also made use of \nustar\ observations of the soft state of \u1630 (observation 40014009001). The inclination constraint also agrees with the lower and upper bounds provided by \cite{Tomsick1998} ($>60^{\circ}$) and \cite{Seifina2014} ($<70^{\circ}$) respectively. However, we note that there is some evidence that the inner disks of BHBs in general, as probed via reflection modeling, may be misaligned with the outer disk, and thus the orbit in general (see, e.g., \citealt{Connors2019_refl}). In addition, we fix the inner disk radius to the ISCO ($R_{\rm in}=R_{\rm ISCO}$), again based upon preliminary fits which revealed its typical value to be very close to $R_{\rm ISCO}$, and poorly constrained. Thus, the reflection properties in models (a) and (b) vary only in the disk ionization $\xi$, the normalization $N_{\rm refl}$, lamppost height $h$, and the disk density $n_{\rm e}$, which varies freely in model (b)---up to $10^{22}~{\rm cm^{-3}}$---but is fixed at low density in model (a). With this carefully curated modeling approach we are able to capture the general differences between low and high density reflection modeling by inspecting the key parameter trends with maximal degrees of freedom in the spectral fits. 

Finally, during preliminary fits with both Models (a and b) we also noticed that the PCA detects the well-known \ion{Fe}{26} absorption feature at $\sim6.9$~keV, associated with an ionized wind, and previously detected in \u1630 \nustar\ spectra during the soft state (see, e.g., \citealt{King2014,Hori2014}). The feature is visible in spectra from the $2002\mbox{--}2004$ outburst at hardness ratios below $\sim0.4$ (see Figure~\ref{fig:rxte_data} for reference). As a result, we apply the {\tt gabs} model to those datasets in which the feature is present, in order to factor out those residuals. We fix the centroid energy of the line to $6.9$~keV, the Gaussian width to $0.01$, and allow only the strength of the line to vary freely. 

\subsubsection{Spectral Fitting Results} \label{subsec:pca_results}
The results of our modeling of the PCA data are summarized visually in Figure~\ref{fig:par_trends}, with numerical parameter values and their confidence limits presented in Tables~\ref{tab:pca_fits_1998} and \ref{tab:pca_fits_2002} in Appendix~\ref{app:pcafits}. The first and most fundamental result to highlight is that the disk density, $n_{\rm e}$, typically trends towards values $>10^{20}~{\rm cm^{-3}}$ across all spectral states. Secondly, there are some clear and significant differences in the properties of the corona when comparing low and high density reflection models. As the model reaches higher densities, lower values of $\Gamma$ and $kT_{\rm e}$ appear to be preferred, particularly during the more intermediate states---this is far clearer in our results for the $1998$ outburst data, whereas in our modeling of $2002\mbox{--}2004$ the systematic differences are much less obvious, which is likely because the model does not fit as well to those observations. Thus it appears that when higher BHB disk densities are assumed (which is arguably more appropriate based on simple BH mass-scaling arguments, see, e.g., \citealt{Garcia2016}), one should systematically expect to derive a cooler corona with a harder IC continuum. We discuss the significance of this result in more detail in Section~\ref{sec:discussion}. 


During the $1998$ outburst, in which the source was caught in a harder state, there are clear differences in the disk properties ($T_{\rm in}$ and $N_{\rm disk}$) between Model (a) and Model (b), as well as a weaker constraint on the coronal height, $h$, allowing a lamppost at almost $100~R_{\rm g}$. The caveat of such statements is that observation 30178-01-01-00 has a low number of counts, and as shown by the reduced $\chi^2$ panel at the bottom of Figure~\ref{fig:par_trends}, the data are over-fit. Nonetheless, it is curious that the introduction of softer reflected emission due to high-density effects appears to be more consistent with a hotter disk with a smaller inner radius. 

We note that Figure~\ref{fig:par_trends} does not show results for the low-density reflection modeling of observations 80117-01-03-00G and 80117-01-07-01 (at hardness ratios of 0.60 and 0.53 respectively). This is because the low-density model (Model a) struggles to fit the data within our parameter setup (with the inclination and inner radius frozen). Rather than show results of the improvements we can make to the modeling with additional free parameters, we choose to show the stricter comparisons across hardness.

Figure~\ref{fig:par_trends} also shows that generally the disk ionization, presented in log units ($\log[\xi/{\rm (erg~cm~s^{-1})}$), is $\sim4$, but in the much softer states reached during the $2002\mbox{--}2004$ outburst it drops two orders of magnitude to $< 2$. This significant decline may not be meaningful because the reflection model dominates the fit over the IC component---thus the constrained value of $\log\xi$ is less reliable, despite the trend coinciding with the drop in coronal flux as expected. In addition, we should expect the bright disk emission to be contributing to the disk ionization, which makes the drop in flux even more suspect. This issue was highlighted in full by \cite{Connors2020}, and is possibly evidence that a IC continuum is an inappropriate description of the irradiating continuum in BHB soft states (due to the superior disk flux with respect to the IC emission). We address this in modeling of the \nustar\ and \swift-XRT data in Section~\ref{sec:nustar_fits}. 

There are not many clear differences in the parameter trends between fits to the $1998$ and $2002\mbox{--}2004$ outbursts. However, one can notice higher disk temperatures during the $2002\mbox{--}2004$ outburst, particularly as the source reached a softer spectral state, where we would expect a higher temperature disk. Constraints on quantities such as the coronal electron temperature, $kT_{\rm e}$, scattering fraction, $f_{\rm sc}$, and coronal height, $h$, are too weak to properly distinguish the differences between outbursts. The electron temperature $kT_{\rm e}$ is best constrained in the intermediate states, where we find good agreement between outbursts, despite the doubling in count rate during the latter outburst in $2002\mbox{--}2004$ (though as Tables~\ref{tab:pca_fits_1998} and \ref{tab:pca_fits_2002} in Appendix~\ref{app:pcafits} show, there are slight numerical differences, with the higher luminosity observations appearing to show a cooler coronal gas).

\subsection{\nustar\ and \swift-XRT modeling} \label{sec:nustar_fits}

We adopt a similar approach when modeling the \nustar\ and \swift-XRT data. Here, however, we move on from comparing low and high density modeling, and instead look to compare high-density reflection from coronal illumination of the disk,  with the reflected returning disk radiation, following the analysis presented in \cite{Connors2020}. In addition, we perform joint spectral modeling, linking key model parameters to better constrain their values across all observations.

\subsubsection{Model Setup} \label{subsec:nustar_setup}

The two reflection models we use are {\tt relconv$\otimes$reflionxHD} and {\tt relxillNS}. The model {\tt relxillNS} was implemented in \cite{Connors2020}, and will be described in detail in a forthcoming paper (Garcia et al., in preparation). It is a variant of the {\tt relxill} consortium of relativistic reflection models \citep{Garcia2014,Dauser2014}, and behaves similarly to {\tt relxill}, the difference being that the irradiating continuum is a single-temperature blackbody spectrum characterized by $kT_{\rm refl}$, as opposed to a cuttoff power law or IC spectrum. 

We use {\tt relxillNS} purely as a consequence of empirical results from modeling, showing that power-law-like reflection proves inadequate during the soft spectral state \citep{Connors2020}. The fundamental driver of this disagreement is the strength of the Compton hump in the reflection component, which is prominent in reflection models which assume an IC irradiating continuum. Reflection from an irradiating blackbody spectrum, on the other hand, naturally results in a weaker Compton hump due to the softer irradiating continuum. The {\tt relxillNS} component acts as a proxy for the reflection spectrum resulting from returning disk radiation illuminating the disk. This is not a model of returning radiation, and the blackbody spectrum we use is a single-temperature one, as opposed to a disk spectrum, so there are strong caveats to using this model. However, it provides us with a way to compare the expected outcome when the disk is reflecting a blackbody-like spectrum as opposed to a power-law-like spectrum. 

\begin{figure}
\centering
\includegraphics[width=\linewidth]{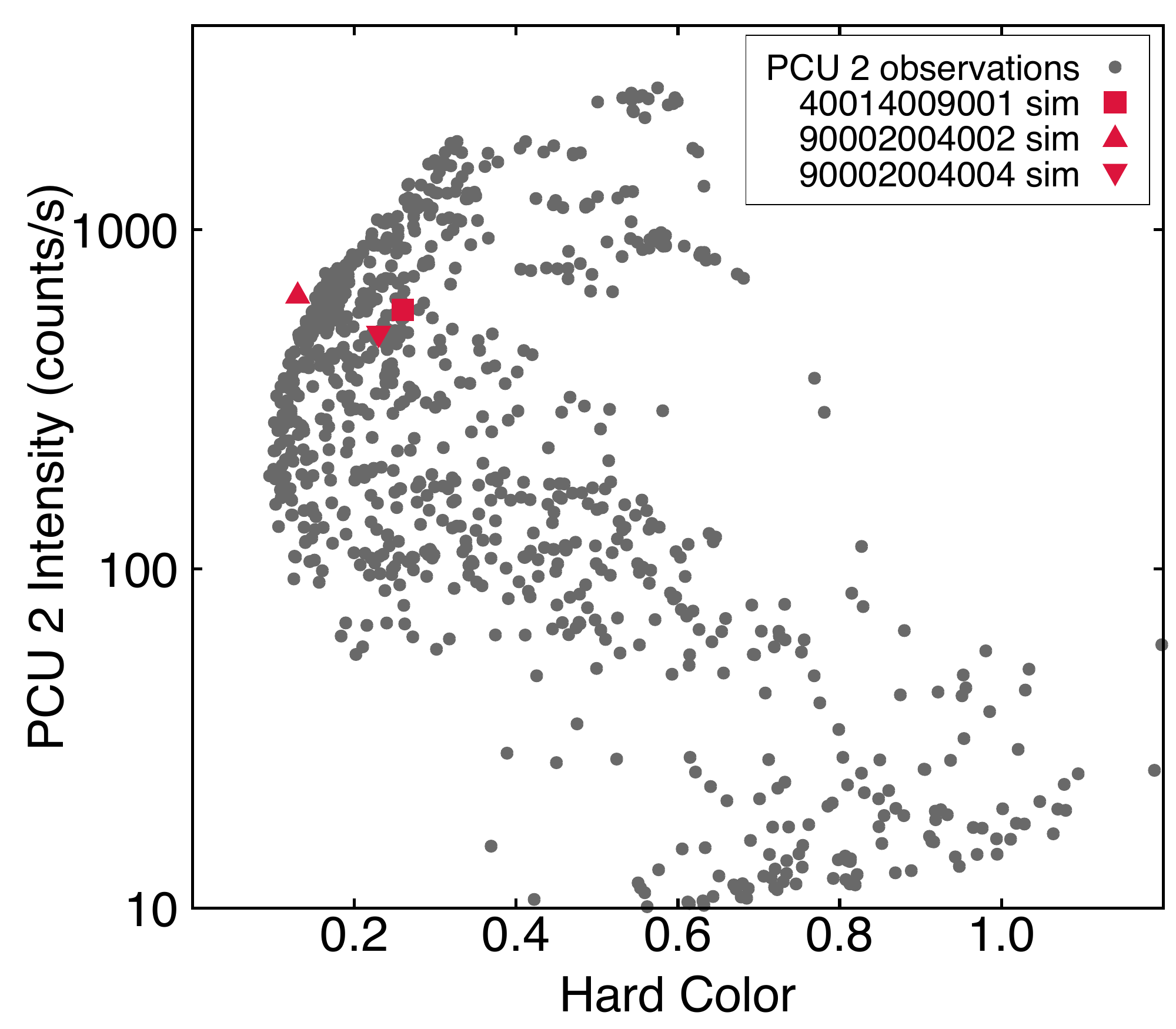}
\caption{HID showing the archival \rxte-PCA (PCU 2) observations of \u1630 with simulated PCU-2 spectra overplotted. The simulated data are generated from a simple model fit to each of the 3 \nustar\ observations of \u1630. The spectral model used is {\tt TBabs(simplcut$\otimes$diskbb)*gabs}, and we assume a PCU~2 detector response corresponding to observation 90410-01-03-00 ($2002\mbox{--}2004$ soft state observation), and an exposure time of 2.5ks. }
\label{fig:simulations}
\end{figure}

This model comparison is suitable given how disk-dominated the source was during the \nustar/\swift\ observations, as shown by simulated data presented in Figure~\ref{fig:simulations}, alongside the superior energy resolution of \nustar\ (a feature which allows for stricter reflection model comparisons), and the soft X-ray coverage provided by \swift-XRT. The simulated PCA observations are based on model fits to the \nustar\ spectra where the assumed model is an absorbed Comptonized disk with an absorption line from an ionized wind: {\tt TBabs(simplcut*diskbb)*gabs}. One can see that all three \nustar/\swift\ observations occurred during the soft state, and basic fits reveal typical disk temperatures of $\sim1.4$~keV.

We described the data reduction and grouping in Section~\ref{sec:nustar_data}. Here we describe the details of the model treatment and fitting procedure. The first model is similar to Model~(b) as presented in Section~\ref{sec:pca_fits}, instead now we require additional calibration corrections to fit the \nustar\ and \swift-XRT spectra simultaneously, as well as the inclusion of dust scattering effects in the model \citep{Kalemci2018}. 

In addition, all the \nustar\ spectra clearly display a strong $\sim6.9$~keV absorption line, commensurate with a previously confirmed ionized disk wind \citep{DiazTrigo2014,King2014}. Given the increased spectral resolution of \nustar\ with respect to \rxte, we replaced our simplistic Gaussian model for the absorption line with the full, self-consistent {\tt xstar} photoionization model \citep{Kallman2001}. We generate a grid of models assuming solar metallicity, and an input blackbody spectrum at a temperature of $1.4$~keV, based on initial absorbed disk$+$powerlaw fits to the simultaneous \nustar\ and \swift-XRT observations, showing a disk inner temperature of $1.4$~keV. We fix the gas density to $n=10^{15}~{\rm cm^{-3}}$, based on typical X-ray wind densities \citep{Miller2008b}. We chose an ionizing luminosity of $10^{38}~{\rm erg~s^{-1}}$, and a turbulent velocity of $1000~{\rm km~s^{-1}}$, based upon the variation in its best fit value identified by \cite{DiazTrigo2014}. Typical wind velocities can fall in the range of $300\mbox{--}3000~{\rm km~s^{-1}}$, based upon rough estimates of the Keplerian velocity range corresponding to the radial launching window allowed for thermally driven winds \citep{Ponti2012}. Our grid then has just three variable parameters, the column density, $N_{\rm H, wind}$, and the ionization, $\log\xi_{\rm wind}$, and the outflow velocity of the wind, $v_{\rm wind}$. 

The models, now (b.2) and (c), are outlined in Table~\ref{tab:models}. We replaced the lamppost version of the relativistic convolution model, {\tt relconvlp}, with {\tt relconv}, which instead parameterizes the emissivity profile of irradiation from $R_{\rm in}$ to $R_{\rm out}$ via the index $q$, where $\epsilon(r)\propto r^{-q}$. This is based on the lack of a lamppost version of the model {\tt relxillNS}, and it therefore allows a more direct comparison of models (b.2) and (c). 

The model {\tt crabcorr} \citep{Steiner2010} applies a correction to the detector response of a given instrument to match the instrument-calibrated normalization and power-law slope of the Crab spectrum \citep{Toor1974}. Thus its model parameters are $N_{\rm CC}$ and $\Delta\Gamma_{\rm CC}$, which renormalize and apply a shift in power law slope respectively. We fix $N_{\rm CC}=1$ and $\Delta\Gamma_{\rm CC}=0$ in FPM A, and allow only $N_{\rm CC}$ to vary for FPM B. In accordance with the results of \cite{Steiner2010}, we apply a $\Delta\Gamma_{\rm CC}=-0.04$ to the \swift-XRT spectra (which are contemperaneous with \nustar\ observations 40014009001 and 90002004004). $N_{\rm CC}$ is then left free to vary as a calibration constant for the \swift-XRT data relative to \nustar. 

\begin{figure*}
\centering
\includegraphics[width=0.49\linewidth]{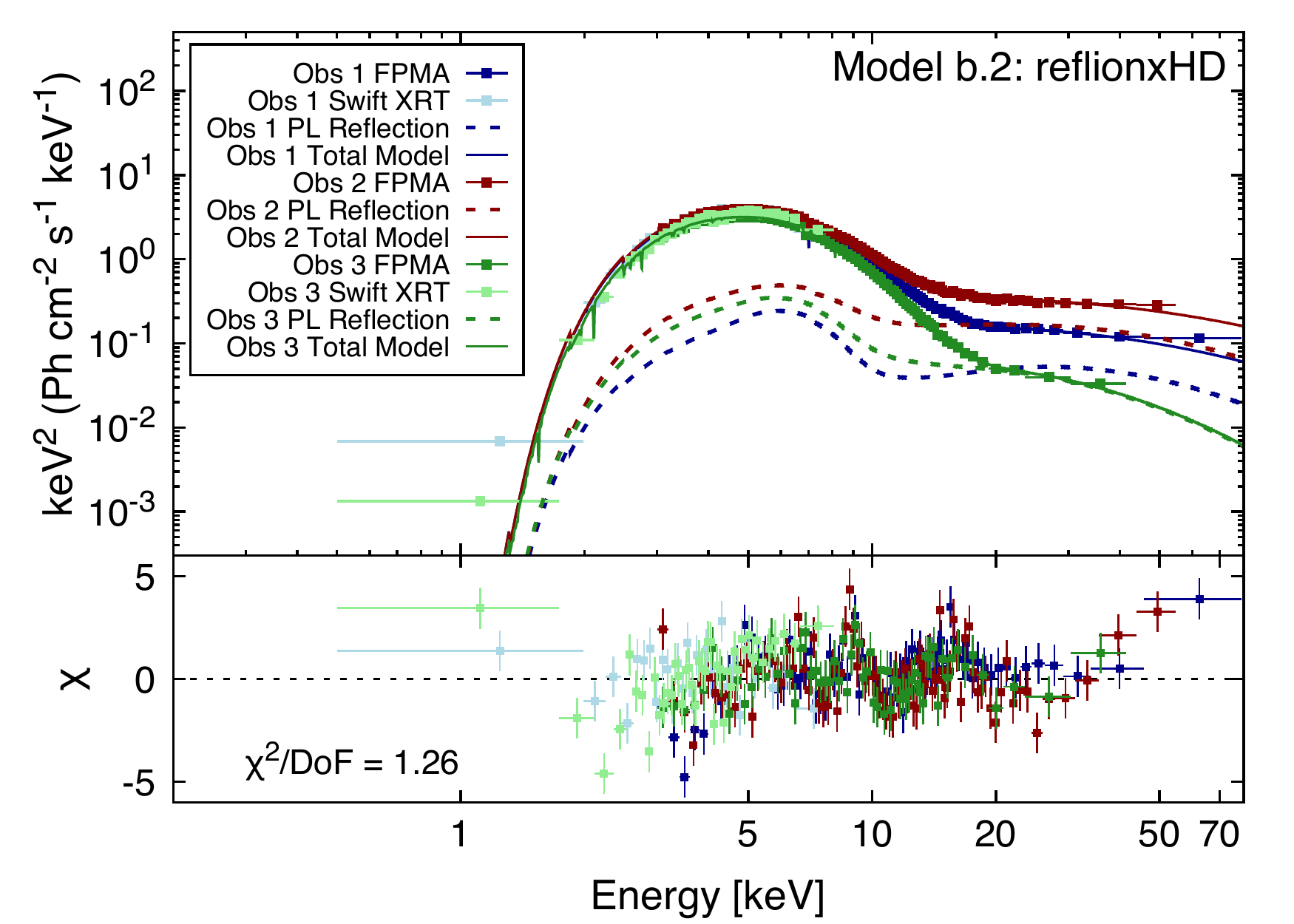}
\includegraphics[width=0.49\linewidth]{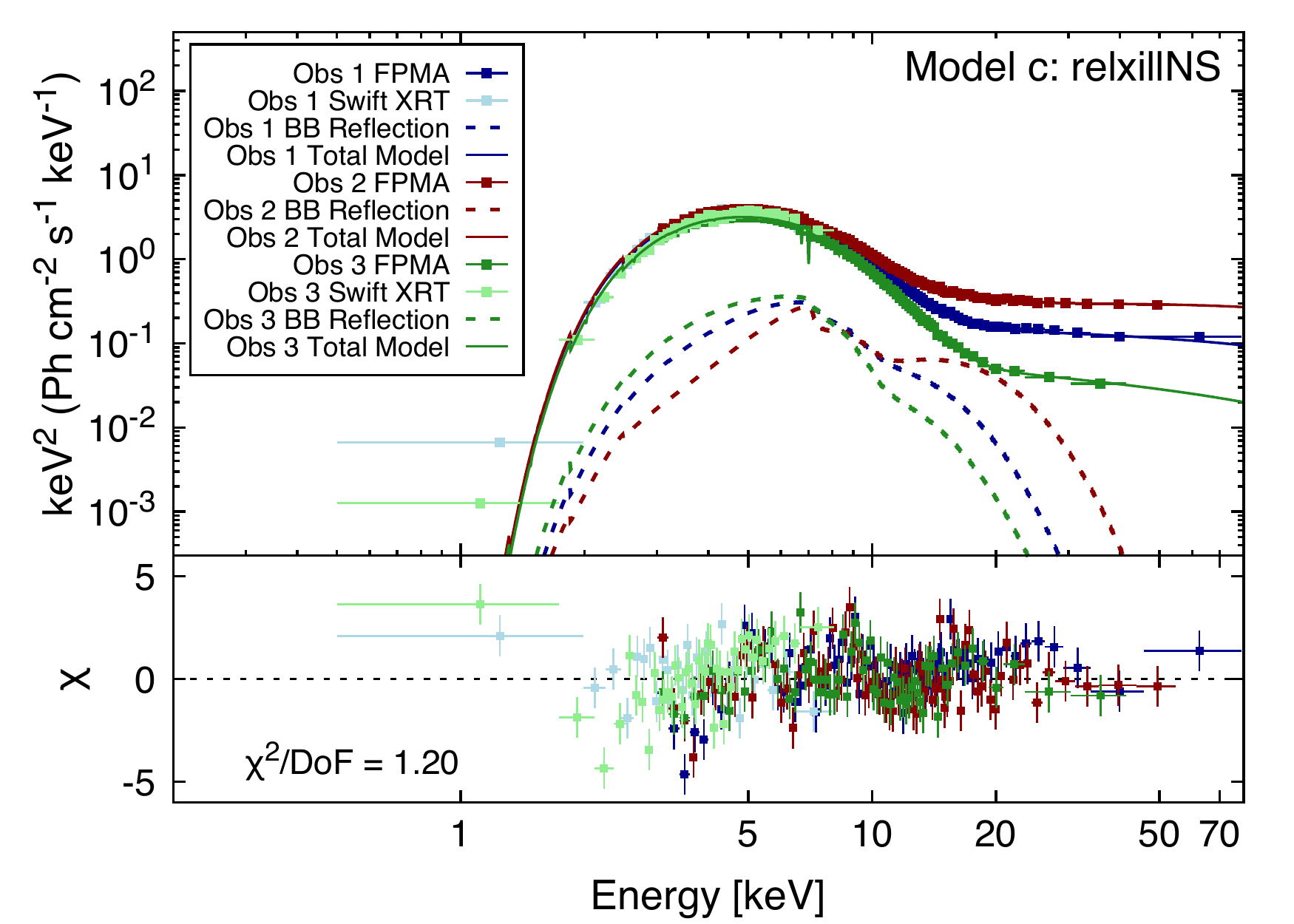}
\caption{Joint reflection modeling of \nustar\ and \swift\-XRT observations of \u1630 taken in 2013 and 2015. The left panel shows the joint fit of Model (b.2) to all three observations, and the right panel shows fits of Model (c)---descriptions are shown in Table~\ref{tab:models}. The {\tt xstar} component has been removed from the reflection components to give clarity to the shape of the reflection spectrum. The bottom section of each panel shows the $\chi$ standardized residuals---(data$-$model)/error. \nustar\ spectra have been rebinned for plotting at a signal-to-noise ratio of $200$ (with a maximum of 5 adjacent bins being combined), and \swift-XRT with a signal-to-noise ratio of $20$ (again with a maximum of 5 adjacent bin combinations).} 
\label{fig:nustar-swift-fits}
\end{figure*}

The model {\tt xscat} \citep{Smith2016} accounts for dust scattering of source photons. For sources with a large hydrogen column density ($N_{\rm H}$)  along the line of sight it is common for a dust scattering halo to form---this is because photons at soft X-ray energies mostly undergo small-angle scattering with dust grains along the line of sight. The effects of dust scattering were explored fully by \cite{Kalemci2018} in the case of \u1630. They found that the bulk of the dust scattering is likely due to a molecular cloud located close to the source ($\sim0.94 D$, where $D$ is the distance to the source). The {\tt xscat} model implementation in Xspec has a few parameters, the most important being $X_{\rm pos}$, the position of the dust scattering halo along the line of sight to the source, $N_{\rm H}$, the hydrogen column density of the scattering source (units of $10^{22}~{\rm cm^{-2}}$), and $R_{\rm ext}$, the radius of the circular extraction region in arcseconds. In all our fits we fix $X_{\rm pos}=0.9$, $R_{\rm ext}=80$" and $R_{\rm ext}=50"$ for the \nustar\ and \swift-XRT spectra respectively. The value of $0.9$ for the position of the scatterer is based on the fact that it is likely that multiple scattering regions exist along the line of sight, despite there being a dominant one located at $X_{\rm pos}=0.94$ \citep{Kalemci2018}. We then tie $N_{\rm H}$ to the line of sight hydrogen column given by the {\tt TBabs} model. This modeling approach limits the different fits to the same level of dust scattering, as it relates to the interstellar gas absorption ($N_{\rm H}$). 

We perform the spectral fits to all three \nustar\ observations (along with the simultaneous \swift-XRT observations) jointly, tying model parameters we have {\it a priori} determined should remain constant. We describe the full parameter setup in the following subsection.

\subsubsection{Parameters} \label{subsec:nustar_parameters}
Given the superior spectral energy resolution of \nustar\ over the PCA, we allow more freedom in the models.  We let the iron abundance, $A_{\rm Fe}$, disk inclination, $i$, BH spin, $a_{\star}$, and absorption column density, $N_{\rm H}$, vary freely, tying them across the three observations such they are assumed to be constant over time. In addition, the disk density, $n_{\rm e}$, is allowed to vary in model (b.2), again with its value tied across the three observations, given all three are at similar count rates, and disk density is dependent on accretion rate. 

As shown by Figure~\ref{fig:simulations}, all three observations are during a soft state, and so we assume the thin accretion disk extends to the ISCO ($R_{\rm in}=R_{\rm ISCO}$). We let $q$ vary freely in fits with Model~b.2, based upon the low-$h$ constraints found in our fits to the PCA data (see Section~\ref{sec:pca_fits}). We fix $q=3$ in our Model~c fits, based on preliminary predictions regarding the shallow emissivity profile of returning disk radiation \citep{Wilkins2020}. Whilst {\tt relxillNS} is not itself a physical model of returning radiation---which should be noted as a significant caveat in our spectral fitting approach---the work of \cite{Wilkins2020} shows that the emissivity profile should be flatter than that of a compact corona irradiating the disk. In Model~c we also separate the temperature of the irradiating blackbody component from the intrinsic inner disk temperature in {\tt diskbb}: $kT_{\rm refl} \neq kT_{\rm in}$. This is based on the inherent uncertainty regarding the relativistic energy shifts experienced by the returning disk photons. We also fix the disk density in {\tt relxillNS} (Model c) to the highest available value the reflection table model affords us, $10^{19}~{\rm cm^{-3}}$. 

We let the {\tt xstar} parameters vary freely for each individual dataset. These are the wind column density, $N_{\rm H, wind}$, gas ionization, $\log\xi_{\rm wind}$, and wind velocity, $v_{\rm wind}$. The column density parameter has limits in our generated {\tt xstar} table given by $5\times10^{21}\mbox{--}5\times10^{23}$, and the ionization has limits given by $2\mbox{--}5$ in log units. The wind velocity is determined by a freely varying redshift parameter on the {\tt xstar} table model. 



\def\oImIXpos{$0.83^{+0.04}_{-0.03}$}
\def\oImINHtbabs{$9.50^{+0.11}_{-0.08}$}
\def\oImIga{$2.39^{+0.03}_{-0.03}$}
\def\oImIfsc{$0.042^{+0.002}_{-0.003}$}
\def\oImIkTe{$300^a$}
\def\oImITin{$1.402^{+0.002}_{-0.003}$}
\def\oImIndisk{$208^{+2}_{-1}$}
\def\oImIq{$>9.8$}
\def\oImIincl{$67^{+1}_{-6}$}
\def\oImIa{$0.987^{+0.002}_{-0.006}$}
\def\oImIlogxi{$2.69^{+0.07}_{-0.07}$}
\def\oImIAfe{$>4.2$}
\def\oImIDensity{$>60$}
\def\oImInrel{$0.61^{+0.05}_{-0.05}$}
\def\oImILineE{$6.97\pm0.02$}
\def\oImISigma{$<110$}
\def\oImIStrength{$0.034\pm0.003$}
\def\oImINHxstar{$10.7^{+0.1}_{-3.8}$}
\def\oImIlogxixstar{$4.17^{+0.03}_{-0.18}$}
\def\oImIvxstar{$2100^{+600}_{-900}$}
\def\oImINccFPMA{$1^a$}
\def\oImINccFPMB{$0.955\pm0.002$}
\def\oImINccSwift{$1.25\pm0.02$}
\def\oImIchi{$1173$}
\def\oImInu{$869$}
\def\oImIchired{$1.35$}

\def\oImIIXpos{$0.751^{+0.030}_{-0.003}$}
\def\oImIINHtbabs{$9.6\pm0.1$}
\def\oImIIga{$2.11^{+0.04}_{-0.07}$}
\def\oImIIfsc{$0.038^{+0.004}_{-0.004}$}
\def\oImIIkTe{$300^a$}
\def\oImIITin{$1.374^{+0.003}_{-0.008}$}
\def\oImIITrefl{$2.01^{+0.12}_{-0.08}$}
\def\oImIIndisk{$230^{+6}_{-3}$}
\def\oImIIq{$3^a$}
\def\oImIIincl{$35^{+3}_{-2}$}
\def\oImIIa{$>0.8$}
\def\oImIIlogxi{$3.47^{+0.03}_{-0.31}$}
\def\oImIIAfe{$>9.3$}
\def\oImIIDensity{$0.1^a$}
\def\oImIInrel{$0.36^{+0.04}_{-0.07}\times10^{-3}$}
\def\oImIILineE{$6.97\pm0.02$}
\def\oImIISigma{$130^{+30}_{-40}$}
\def\oImIIStrength{$0.046\pm0.005$}
\def\oImIINHxstar{$10.8^{+0.2}_{-3.1}$}
\def\oImIIlogxixstar{$4.09^{+0.04}_{-0.04}$}
\def\oImIIvxstar{$2100^{+600}_{-600}$}
\def\oImIINccFPMA{$1^a$}
\def\oImIINccFPMB{$0.955\pm0.002$}
\def\oImIINccSwift{$1.26\pm0.02$}
\def\oImIIchi{$1137$}
\def\oImIInu{$870$}
\def\oImIIchired{$1.31$}

\def\oIImIXpos{$0.49^{+0.22}_{-0.04}$}
\def\oIImINHtbabs{$9.6\pm0.1$}
\def\oIImIga{$2.21^{+0.02}_{-0.02}$}
\def\oIImIfsc{$0.053^{+0.003}_{-0.006}$}
\def\oIImIkTe{$300^a$}
\def\oIImITin{$1.411^{+0.001}_{-0.003}$}
\def\oIImIndisk{$204^{+8}_{-1}$}
\def\oIImIq{$9.5^{+0.4}_{-0.3}$}
\def\oIImIincl{$21^{+8}_{-18}$}
\def\oIImIa{$<0.6$}
\def\oIImIlogxi{$3.27^{+0.08}_{-0.17}$}
\def\oIImIAfe{$<0.53$}
\def\oIImIDensity{$10^{+2}_{-6}$}
\def\oIImInrel{$0.68^{+0.02}_{-0.03}$}
\def\oIImILineE{$6.97\pm0.02$}
\def\oIImISigma{$<100$}
\def\oIImIStrength{$0.019^{+0.0}_{-0.003}$}
\def\oIImINHxstar{$10.7^{+0.1}_{-0.3}$}
\def\oIImIlogxixstar{$4.43^{+0.03}_{-0.09}$}
\def\oIImIvxstar{$3000^{+900}_{-1200}$}
\def\oIImINccFPMA{$1.149^{+0.009}_{-0.006}$}
\def\oIImINccFPMB{$1.015^{+0.001}_{-0.001}$}
\def\oIImINccSwift{\nodata}
\def\oIImIchi{$989$}
\def\oIImInu{$820$}
\def\oIImIchired{$1.21$}

\def\oIImIINHxscat{$>30$}
\def\oIImIIXpos{$0.94^a$}
\def\oIImIINHtbabs{$9.77^{+0.08}_{-0.10}$}
\def\oIImIIga{$1.94^{+0.02}_{-0.04}$}
\def\oIImIIfsc{$0.055^{+0.003}_{-0.004}$}
\def\oIImIIkTe{$300^a$}
\def\oIImIITin{$1.404^{+0.005}_{-0.004}$}
\def\oIImIITrefl{$>2.93$}
\def\oIImIIndisk{$229.5^{+2.1}_{-0.9}$}
\def\oIImIIq{$3^a$}
\def\oIImIIincl{$43^{+3}_{-4}$}
\def\oIImIIa{$>0.86$}
\def\oIImIIlogxi{$3.01^{+0.03}_{-0.08}$}
\def\oIImIIAfe{$>8$}
\def\oIImIIDensity{$0.1^a$}
\def\oIImIInrel{$0.34^{+0.04}_{-0.03}\times10^{-3}$}
\def\oIImIILineE{$7.04^{+0.05}_{-0.04}$}
\def\oIImIISigma{$0.23\pm0.06$}
\def\oIImIIStrength{$0.050^{+0.013}_{-0.007}$}
\def\oIImIINHxstar{$1.9^{+0.6}_{-0.4}$}
\def\oIImIIlogxixstar{$3.3^{+0.2}_{-0.2}$}
\def\oIImIIvxstar{$12000^{+1000}_{-2000}$}
\def\oIImIINccFPMA{$1.015\pm0.001$}
\def\oIImIINccFPMB{$1.015\pm0.001$}
\def\oIImIINccSwift{\nodata}
\def\oIImIIchi{$933$}
\def\oIImIInu{$821$}
\def\oIImIIchired{$1.14$}

\def\oIIImIXpos{$0.9^a$}
\def\oIIImINHtbabs{$10.1^{+0.1}_{-0.2}$}
\def\oIIImIga{$2.91^{+0.03}_{-0.04}$}
\def\oIIImIfsc{$<0.02$}
\def\oIIImIkTe{$300^a$}
\def\oIIImITin{$1.397^{+0.002}_{-0.002}$}
\def\oIIImIndisk{$175^{+1}_{-2}$}
\def\oIIImIq{$>9.5$}
\def\oIIImIincl{$65^{+2}_{-5}$}
\def\oIIImIa{$0.978^{+0.006}_{-0.024}$}
\def\oIIImIlogxi{$2.90^{+0.02}_{-0.05}$}
\def\oIIImIAfe{$>4$}
\def\oIIImIDensity{$>20$}
\def\oIIImInrel{$<0.7$}
\def\oIIImInxil{$80^{+10}_{-10}$}
\def\oIIImILineE{$6.92\pm0.01$}
\def\oIIImISigma{$0.10\pm0.01$}
\def\oIIImIStrength{$0.047^{+0.004}_{-0.003}$}
\def\oIIImINHxstar{$2.9^{+0.1}_{-0.1}$}
\def\oIIImIlogxixstar{$2.57^{+0.05}_{-0.06}$}
\def\oIIImIvxstar{$14400^{+300}_{-600}$}
\def\oIIImINccFPMA{$1.003^{+0.010}_{-0.005}$}
\def\oIIImINccFPMB{$1.007^{+0.001}_{-0.002}$}
\def\oIIImINccSwift{$0.823\pm0.009$}
\def\oIIImIchi{$911$}
\def\oIIImInu{$752$}
\def\oIIImIchired{$1.21$}

\def\oIIImIIXpos{$0.9^a$}
\def\oIIImIINHtbabs{$10.4\pm0.1$}
\def\oIIImIIga{$2.4\pm0.1$}
\def\oIIImIIfsc{$0.022^{+0.004}_{-0.004}$}
\def\oIIImIIkTe{$300^a$}
\def\oIIImIITin{$1.398^{+0.008}_{-0.008}$}
\def\oIIImIITrefl{$1.6^{+0.1}_{-0.1}$}
\def\oIIImIIndisk{$173^{+8}_{-3}$}
\def\oIIImIIq{$3^a$}
\def\oIIImIIincl{$47\pm1$}
\def\oIIImIIa{$>0.88$}
\def\oIIImIIlogxi{$3.2^{+0.3}_{-0.1}$}
\def\oIIImIIAfe{$>9.3$}
\def\oIIImIIDensity{$0.1^a$}
\def\oIIImIInrel{$0.46^{+0.04}_{-0.03}\times10^{-3}$}
\def\oIIImIILineE{$6.959^{+0.02}_{-0.008}$}
\def\oIIImIISigma{$0.22\pm0.03$}
\def\oIIImIIStrength{$0.10\pm0.01$}
\def\oIIImIINHxstar{$10.7^{+0.2}_{-1.3}$}
\def\oIIImIIlogxixstar{$3.90^{+0.04}_{-0.06}$}
\def\oIIImIIvxstar{$1800^{+600}_{-600}$}
\def\oIIImIINccFPMA{$0.80\pm0.06$}
\def\oIIImIINccFPMB{$1.007\pm0.002$}
\def\oIIImIINccSwift{$0.824\pm0.009$}
\def\oIIImIIchi{$843$}
\def\oIIImIInu{$753$}
\def\oIIImIIchired{$1.12$}
\def\oIIImIIedgeE{$4.33^{+0.06}_{-0.06}$}
\def\oIIImIIMaxTau{$0.033^{+0.003}_{-0.003}$}

\def\JointmIa{$0.989^{+0.001}_{-0.002}$}
\def\JointmIincl{$69.7^{+0.4}_{-0.9}$}
\def\JointmIDensity{$>80$}
\def\JointmINH{$9.49^{+0.04}_{-0.04}$}
\def\JointmIAfe{$>4.9$}
\def\JointmIchi{$3080$}
\def\JointmInu{$2445$}
\def\JointmIchired{$1.26$}
\def\JointmIIa{$0.85^{+0.07}_{-0.07}$}
\def\JointmIIincl{$37^{+1}_{-2}$}
\def\JointmIIDensity{$0.1^a$}
\def\JointmIINH{$9.64^{+0.07}_{-0.06}$}
\def\JointmIIAfe{$>9.2$}
\def\JointmIIchi{$2938$}
\def\JointmIInu{$2446$}
\def\JointmIIchired{$1.20$}

\begin{deluxetable*}{lcccccc}
\tabletypesize{\footnotesize}
\tablecaption{Maximum likelihood estimates of all parameters in spectral fitting of \nustar/\swift-XRT observations of \u1630, comparing models b.2 and c. \label{tab:nustar-params}}
\tablecolumns{7}
\tablehead{
\colhead{Parameter} & 
\multicolumn{3}{c}{Model (b.2)} &
\multicolumn{2}{c}{Model (c)} \\
 &
\colhead{Obs 1} & 
\colhead{Obs 2} & 
\colhead{Obs 3} & 
\colhead{Obs 1} & 
\colhead{Obs 2} &
\colhead{Obs 3} 
}
\startdata
$N_{\rm CC, FPMA}$ & & & \multicolumn{2}{c}{$1$} & & \\
$\Delta\Gamma_{\rm CC, FPMA}$ & & & \multicolumn{2}{c}{$0$} & & \\
$\Delta\Gamma_{\rm CC, FPMB}$ & & & \multicolumn{2}{c}{$0$} & & \\
$\Delta\Gamma_{\rm CC, \swift}$ & {$-0.04$} & \nodata & {$-0.04$} & {$-0.04$} & \nodata & {$-0.04$} \\
$R_{\rm ext, \nustar}$ & & &  \multicolumn{2}{c}{$80''$} &  & \\
$R_{\rm ext, \swift}$ & {$50''$} & \nodata & {$50''$} & {$50''$} & \nodata & {$50"$} \\
$X_{\rm pos, dust}$ & & & \multicolumn{2}{c}{$0.9$} & & \\
\hline
$kT_{\rm e}$~[keV] & & & \multicolumn{2}{c}{$300$} & & \\
$R_{\rm in}~[R_{\rm ISCO}]$ &  & & \multicolumn{2}{c}{$1$} & & \\
$N_{\rm H}~[10^{22}~{\rm cm^{-2}}]$ & \multicolumn{3}{c}\JointmINH & \multicolumn{3}{c}\JointmIINH \\
 $a_{\star}$ & \multicolumn{3}{c}\JointmIa & \multicolumn{3}{c}\JointmIIa  \\
 $i$~[$^\circ$] & \multicolumn{3}{c}\JointmIincl & \multicolumn{3}{c}\JointmIIincl   \\
 $A_{\rm Fe}$~[Solar] & \multicolumn{3}{c}\JointmIAfe & \multicolumn{3}{c}\JointmIIAfe   \\
 $n_{\rm e}~{\rm [10^{20}~cm^{-3}]}$ & \multicolumn{3}{c}\JointmIDensity & \multicolumn{3}{c}\JointmIIDensity  \\
\hline
$\Gamma$ & \oImIga\ & \oIImIga\ & \oIIImIga\ & \oImIIga\ & \oIImIIga\ & \oIIImIIga\   \\
$f_{\rm sc}$ & \oImIfsc\ & \oIImIfsc & \oIIImIfsc\ & \oImIIfsc\ & \oIImIIfsc\ & \oIIImIIfsc\ \\
$kT_{\rm in}$~[keV] & \oImITin & \oIImITin\ & \oIIImITin\ & \oImIITin\ & \oIImIITin\ & \oIIImIITin\ \\
$kT_{\rm refl}$~[keV] & \nodata & \nodata\ & \nodata & \oImIITrefl\ & \oIImIITrefl\ & \oIIImIITrefl\ \\
$N_{\rm disk}$ & \oImIndisk & \oIImIndisk\ & \oIIImIndisk\ & \oImIIndisk\ & \oIImIIndisk\ & \oIIImIIndisk\  \\
$q$ & \oImIq & \oIImIq\ & \oIIImIq\ & \oImIIq\ & \oIImIIq\ & \oIIImIIq\  \\
$\log{\xi}$~[${\rm erg~cm~s^{-1}}$] & \oImIlogxi & \oIImIlogxi\ & \oIIImIlogxi\ & \oImIIlogxi\ & \oIImIIlogxi\ & \oIIImIIlogxi\  \\
$N_{\rm refl}$ & \oImInrel & \oIImInrel\ & \oIIImInrel\ & \oImIInrel\ & \oIImIInrel\ & \oIIImIInrel\  \\
$N_{\rm H, wind}~[10^{22}~{\rm cm^{-2}}]$ & \oImINHxstar\ & \oIImINHxstar\ & \oIIImINHxstar\ & \oImIINHxstar\ & \oIImIINHxstar\ & \oIIImIINHxstar\ \\
$\log\xi_{\rm wind}$~[${\rm erg~cm~s^{-1}}$] & \oImIlogxixstar\ & \oIImIlogxixstar\ & \oIIImIlogxixstar\ & \oImIIlogxixstar\ & \oIImIIlogxixstar\ & \oIIImIIlogxixstar\ \\
$v_{\rm wind}$~[${\rm km~s^{-1}}$] & \oImIvxstar\ & \oIImIvxstar\ & \oIIImIvxstar\ & \oImIIvxstar\ & \oIImIIvxstar\ & \oIIImIIvxstar\ \\
$N_{\rm CC, FPMB}$ & \oImINccFPMB\ & \oIImINccFPMB\ & \oIIImINccFPMB\ & \oImIINccFPMB\ & \oIImIINccFPMB\ & \oIIImIINccFPMB\ \\
$N_{\rm CC, \swift}$ & \oImINccSwift\ & \oIImINccSwift\ & \oIIImINccSwift\ & \oImIINccSwift\ & \oIImIINccSwift\ & \oIIImIINccSwift\ \\
\hline
 $\chi^2$ & \multicolumn{3}{c}\JointmIchi\ & \multicolumn{3}{c}\JointmIIchi\   \\
 $\nu$ & \multicolumn{3}{c}\JointmInu\ & \multicolumn{3}{c}\JointmIInu\  \\
 $\chi_{\nu}^2$ & \multicolumn{3}{c}\JointmIchired\ & \multicolumn{3}{c}\JointmIIchired\  \\
\enddata
\tablecomments{Model (b.2): {\tt crabcorr*xscat*TBabs(simplcut$\otimes$diskbb + relconv$\otimes$reflionxHD)*xstar}. Model (c): {\tt crabcorr*xscat*TBabs(simplcut$\otimes$diskbb + relxillNS)*xstar}. {\tt xstar} represents the ionized wind, where $N_{\rm H, wind}$ is the column density of the wind, $\log\xi_{\rm wind}$ is the wind ionization, and $v_{\rm wind}$ is its outflow velocity. $N_{\rm CC}$ and $\Delta\Gamma_{\rm CC}$ are the normalization and photon index shifts in the component {\tt crabcorr}, shown in the table for each instrument. $R_{\rm ext}$ is the aperture size assumed to calculate dust scattering in the {\tt xscat} component, and $X_{\rm pos, dust}$ is the fractional position of the dust scattering halo, where $1$ corresponds to the source location. The disk normalization is given by $N_{\rm disk} = (R_{\rm in}/\kappa^2 D_{10})^2\cos i$, where $R_{\rm in}$ is the apparent inner disk in km, $D_{10}$ is the distance to the source in units of 10~kpc, $i$ is the disk inclination, and $\kappa$ is the color correction factor. The total $\chi^2$ is shown for each fit, along with the degrees of freedom, $\nu$, and the reduced $\chi^2$, $\chi^2_{\nu}=\chi^2/\nu$. The ionization, $\log\xi$, is given by $4\pi F_{\rm irr}/n_{\rm e}$, where $F_{\rm irr}$ is the ionizing flux, and $n_{\rm e}$ is the gas density. The normalization definition of the reflection models, given by $N_{\rm refl}$, is such that the integrated energy flux from $0.1\mbox{--}1000$~keV is equal to $10^{20}n_{\rm e}\xi/4\pi$. We note however that the density in {\tt relxillNS} (Model c) fits is fixed at $10^{19}~{\rm cm^{-3}}$, leading to systematically lower $N_{\rm refl}$ constraints than found via Model b.2 fits with {\tt reflionxHD}.} All other parameters are as described in the text.
\tablenotetext{a} {Frozen parameter}
\end{deluxetable*}

\subsubsection{Spectral Fitting Results} \label{subsec:nustar_results}
The resulting spectral fits are shown in Figure~\ref{fig:nustar-swift-fits}, and corresponding parameters with 90\% confidence limits are shown in Table~\ref{tab:nustar-params}. The key result to highlight is that when modeling all three observations Model (c) performs slightly better than Model (b.2). The superiority of reflected returning radiation to fit the broadband X-ray spectra is clear both from the lower total $\chi^2$ for one additional degree of freedom, and the patterns in residuals shown in Figure~\ref{fig:nustar-swift-fits}---Model b.2 underfits the high energies and introduces curvature around $10\mbox{--}20$~keV. Thus our results make a strong case for returning radiation as a contributor to the observed reflection spectrum in BHB soft states \citep{Connors2020}. 

The hydrogen column density is consistently within the range of $9.5\mbox{--}9.7\times10^{22}~{\rm cm^{-2}}$, in agreement with recent high-resolution spectroscopic observations with {\it Chandra} \citep{Gatuzz2019}. The higher value ($14\times10^{22}~{\rm cm^{-3}}$) assumed in our fits to the PCA data (see Section~\ref{sec:pca_fits}) may reflect the increased effects of dust scattering due to the large aperture size of the PCA detectors. 

We find that the disk density (when applying Model b.2) exceeds $9\times10^{21}~{\rm cm^{-3}}$, which is consistent with our findings from modeling of PCA data shown in Section~\ref{sec:pca_fits}. However, inspection of Figure~\ref{fig:nustar-swift-fits} also shows that given the source is in the soft state, reflection from illumination of the disk by the corona ({\tt relconv$\otimes$reflionxHD}) tends to over-fit the power law component in the data, introducing an undesirable curvature in the residuals. Indeed the parameter constraints in Table~\ref{tab:nustar-params} shows this quantitatively in the systematic difference in coronal parameters between Models (b.2) and (c): Model c retrieves similar values of $f_{\rm sc}$, but significantly lower values of $\Gamma$, or in other words a much harder power law component. 

The BH spin is high in both Model b.2 and Model c fits, with $a_{\star}=0.989^{+0.001}_{-0.002}$ and $a_{\star}=0.85\pm0.07$ respectively. The Model b.2 constraint on BH spin thus agrees fairly well with those found by \cite{King2014} in their analysis of the 2013 observations of \u1630, but the Model c constraint is considerably lower, albeit remaining a high BH spin. This is encouraging given that the models differ slightly, with \cite{King2014} adopting the model {\tt refbhb} \citep{Ross2007} which includes the transmitted emission from internal disk dissipation. 


The Model b.2 inclination constraint agrees well with binary orbit inclination limits ($60^{\circ}<i<70^{\circ}$; \citealt{Tomsick1998,Seifina2014}), as well as with modeling by \cite{King2014}. However, we find a systematic difference in the inclination obtained from Model c (returning radiation) fits, with $i=37^{+1}_{-2}~{\rm deg}$. We note that {\tt relxillNS} is a very preliminary model lacking physical consistency, and so we cannot derive much meaning from this difference in inclination. One can speculate that a contrast in inclination with respect to power-law reflection modeling may be expected since there is additional curvature around the Fe line region, as well as a weaker Compton hump---once the reflection line and continuum adjust significantly, one may expect the inclination to re-adjust to fit the data. We refrain from going beyond this basic explanation, in an effort to avoid speculation on the physical interpretation.

Our ionizing wind parameter constraints reveal a similar dichotomy outlined by \cite{King2014}, between ionization state and blueshift of Fe absorption lines. We find that some of our Model/observation combinations (see Table~\ref{tab:nustar-params}) prefer wind velocities on the order of $\sim2000~{\rm km~s^{-1}}$ with higher wind ionization levels, $\log\xi_{\rm wind}/[{\rm erg~cm~s^{-1}}]\sim4$. Others prefer an ultra-fast outflow with $v_{\rm wind}>10,000~{\rm km~s^{-1}}$, and ionizations closer to $\log\xi_{\rm wind}/[{\rm erg~cm~s^{-1}}]\sim3$. Whilst the absolute values differ from the results of \cite{King2014}, the trends are similar. However, high-resolution spectroscopy provided the best constraints on the disk wind, revealing a relatively low velocity, high ionization, and high column density \citep{Kubota2007,DiazTrigo2014}. This is indeed what we find for four of the six model fits. There is no clear and obvious distinction between Models b.2 and c in terms of this dichotomy in the wind properties, so we suggest that the dichotomy is predominantly due to the limiting resolution of the \nustar\ spectra compared to, for example, {\it Chandra} high energy transmission grating (HETG) spectra and {\it XMM-Newton} spectra. This explains why we \cite{King2014} derived a similar dichotomy. 


\section{Discussion and Conclusions} \label{sec:discussion}
We have presented detailed reflection modeling of \rxte-PCA (Section~\ref{sec:pca_fits}), \nustar\ and \swift-XRT (Section~\ref{sec:nustar_fits}) observations of \u1630, exploring a few different physical models for reflection: IC irradiation of both a low and high density accretion disk, and se-lf-irradiation of the disk by returning disk blackbody radiation.

 We found that when fitting a reflection model with variable disk density to the PCA observations of \u1630 one finds densities on the order of $10^{20}~{\rm cm^{-3}}$. We also showed that when compared to reflection modeling with a fixed, low density reflector ($n_{\rm e}=10^{15}~{\rm cm^{-3}}$), the higher density model leads to harder IC spectra with a lower electron temperature---i.e., a cooler Comptonizing corona with a larger optical depth. 

\subsection{Checking Modeling Consistencies} \label{subsec:consistencies}
We can gain some perspective on the consistency of our PCA modeling results via basic calculations of the expected disk ionization as a function of the irradiating flux, as given by the observed coronal IC flux. With this approach, the expected ionization is given by $\xi=4\pi F_{\rm irr}/n_{\rm e}$. The irradiating flux, $F_{\rm irr}$, is a function of the disk-corona geometry. The geometry assumed in our modeling is a lamppost corona at some height $h$, and we assumed a disk extending to the ISCO ($R_{\rm in}=R_{\rm ISCO}$). Then the ionization depends on the disk density, $n_{\rm e}$, another of our model parameters. Thus this simple expression allows us to confirm whether or not the modeling is consistent with our assumption that the disk is close to the ISCO, as well as place limits on the degree of disk truncation.

 Given the degeneracy between the reflected emission and the direct observed coronal emission in the spectral fits, we instead adopt the total $0.1\mbox{--}100$~keV unabsorbed disk flux, and multiply by the upper bound $f_{\rm sc}$ value as an approximation of the observed coronal flux ($F_{\rm IC}$). Assuming an isotropically emitting source, the observed coronal luminosity, $L_{\rm irr}$, is given by $4\pi D^2 F_{\rm IC}$, $D=10.5~{\rm kpc}$. The intrinsic coronal luminosity is then calculated by incorporating relativistic corrections. We then perform full GR ray tracing to calculate the irradiating flux on the disk, assuming a BH mass of $10~M_{\odot}$, and BH spin parameter $a_{\star}=0.998$. Ray tracing is performed using the {\tt relxill}\ model in the lamppost geometry \citep{Dauser2013}.
 
 The disk ionization parameter, $\log\xi$, in the {\tt reflionxHD} model, is an average disk ionization (it has a fixed value with disk radius). Therefore, the observed disk ionization we determine from fitting the model to data is an average value, while we would expect the actual ionization on the disk to change with the emissivity profile. In order to get an estimate of allowed ionizations that our model predicts, we also calculate the ionization for the radii where the observer sees most of the flux. To be conservative, we use the radii enclosing the area emitting from 10\% to 90\% of the total observed flux, as counted from the edge of the disk. 
 
 Additionally, for the upper ionization limit, we use the lower bound of the lamp post height, and the lower bound of the disk density, $n_{\rm e}$, such that we are predicting an absolute upper limit on the irradiating flux and disk ionization state for our model fits. The values are taken from Table~\ref{tab:pca_fits_1998}. Similary, we also calculate the lower ionization limit, taking respectively the upper limit on the height and disk density. 
\begin{figure}
\centering
\includegraphics[width=\linewidth]{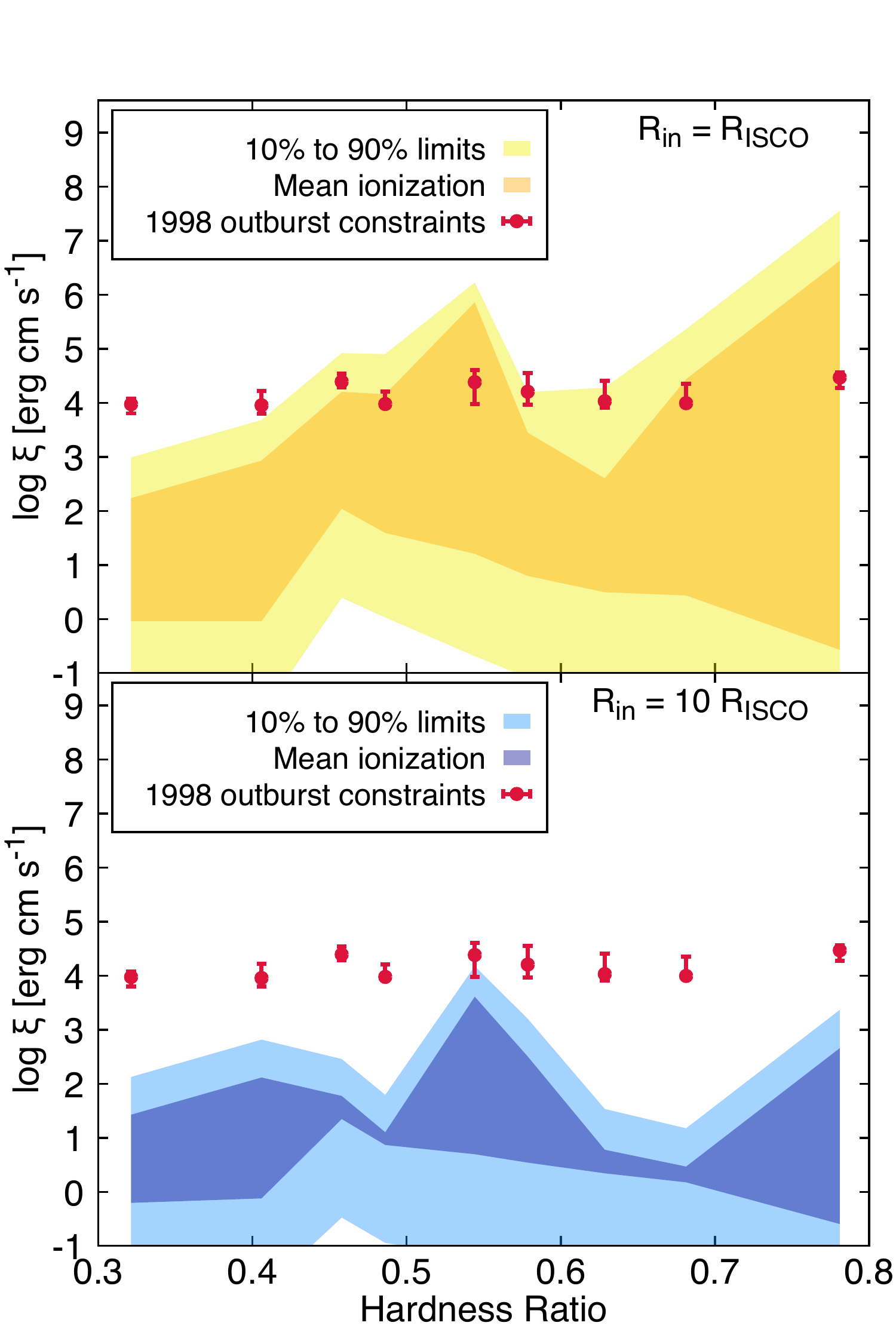}
\caption{Disk ionization, $\log\xi$, as a function of hardness ratio, showing both the measured values from reflection modeling of the $1998$ outburst data (red), and the predicted value given by $\xi=4\pi F_{\rm irr}/n_{\rm e}$ (yellow and blue shaded regions). The top panel shows our predictions for a disk extending to the ISCO, and the bottom panel instead for a disk truncated out to $10~R_{\rm ISCO}$. Ionizations are calculated for the upper and lower 90\% confidence limits on $h$ and $n_{\rm e}$, the lamppost height and disk density (see Table~\ref{tab:pca_fits_1998}). The light shaded regions (yellow and blue) show the maximal observed range of disk ionization. This is given by the ionization at the location at which 10\% (upper bound) of the total irradiating flux has struck the disk, and then 90\% (lower bound). The mean observed ionization (the ionization at the disk radius whereby the disk receives its median flux) is then shown in darker shaded regions (orange and dark blue), again folding in the uncertainty on $h$ and $n_{\rm e}$.  }
\label{fig:ionization}
\end{figure}

We show this range and how it compares to the constrained values of $\log\xi$ from the 1998 outburst spectral fits in Figure~\ref{fig:ionization}. If the disk sits at the ISCO, particularly during the hardest spectral observation in our sample, the lack of constraint on $h$ means we derive a potential peak disk ionization which exceeds the measured value by 3 orders of magnitude. However, if either the corona is less compact ($h$ is large), or the disk is slightly truncated (out to just $10~R_{\rm ISCO}$), then the contribution to the ionization of the disk from coronal illumination sits below the measured value for all values of $h$. This is actually true of our fits to all 9 spectra, as displayed by the blue shaded region in the bottom panel of Figure~\ref{fig:ionization}. 

\begin{figure}
\centering
\includegraphics[width=\linewidth]{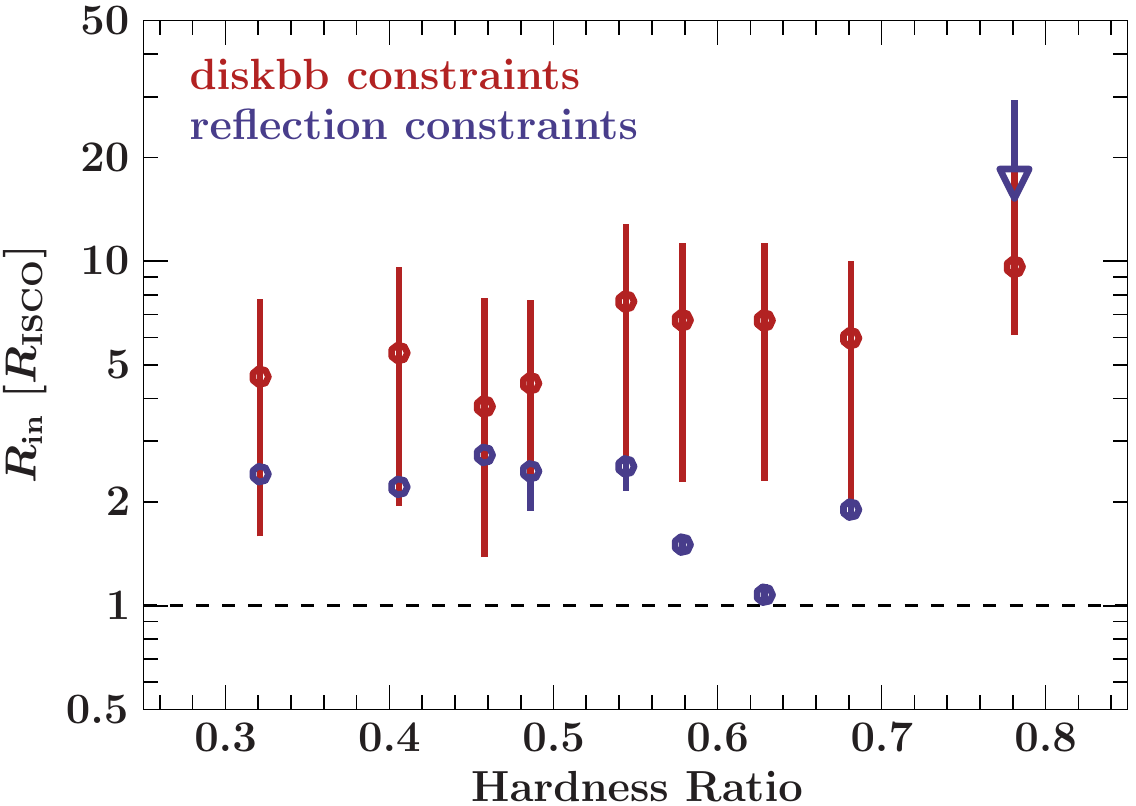}
\caption{Constraints on the inner disk radius, $R_{\rm in}$, from both the multi-temperature disk blackbody component ({\tt diskbb}; red) and the relativistic reflection component ({\tt relconv$\otimes$reflionxHD}; blue) as given by model (b) fits to the \rxte-PCA 1998 outburst data. }
\label{fig:rin_diskbb}
\end{figure}

 In softer states, when the coronal flux has decreased significantly, we underestimate the constrained disk ionization. This disagreement may be occurring for several reasons. However, it is most likely a combination of power-law-like reflection being an inappropriate model for the soft state \citep{Connors2020}, as well as a result of neglecting the disk flux, ie., self-irradiation and heating from the inner regions of the disk, which will contribute significantly to ionizing the upper layers of the disk. Without the data quality to test these degeneracies further, we cannot make any strong statements about inner radius constraints. This simple cross-check instead displays the consistency of key scalings such as disk density and ionizing flux. {\it Perhaps most importantly, these ionization constraints support the implementation of high-density reflection models in X-ray spectral modeling of BHs}. 
 
 In addition to utilizing the disk ionization state as a benchmark for the degree of truncation, one can derive the inner radius directly from the {\tt diskbb} normalization parameter, $N_{\rm disk}$. We performed the same fits presented in Section~\ref{sec:pca_fits} with two key differences: we fixed the disk density to $n_{\rm e}=10^{20}~{\rm cm^{-3}}$, and let the inner disk radius, $R_{\rm in}$, vary freely. This allows us to make a direct comparison between the {\tt diskbb} normalization constraint on $R_{\rm in}$, as well as constraints from the relativistic reflection component given by {\tt relconv$\otimes$reflionxHD}. The inner disk radius can be derived from the disk normalization parameter as outlined by \cite{Mitsuda1984}, whereby $R_{\rm in}~{\rm [km]}=\kappa^2 D_{10} (N_{\rm disk} / \cos i)^{1/2}$, where $\kappa$ is the color correction factor \citep{kubota1998}, $D_{10}$ is the distance to the source in units of $10~{\rm kpc}$, and $i$ is the disk inclination. We assume a fixed disk inclination of $64^{\circ}$ as in the spectral fits presented in Section~\ref{sec:pca_fits}, and we adopt the best fit $N_{\rm disk}$ values with their uncertainties. We then assume a nominal value for the color correction of $\kappa=1.5$, and allow for values between $1$ and $2$. We then propagate the uncertainties on both $N_{\rm disk}$ and $\kappa$ to calculate $R_{\rm in}$ and its uncertainties. Figure~\ref{fig:rin_diskbb} shows the comparison of disk and reflection constraints on $R_{\rm in}$ in units of $R_{\rm ISCO}$.   
 
 The large uncertainties on $R_{\rm in}$ predictions derived from the $N_{\rm disk}$ constraints are largely due to the uncertainty on the color correction factor ($1\le\kappa\le2$). With these uncertainties incorporated into the predicted values, just two of the nine disk and reflection $R_{\rm in}$ predictions disagree, and only on the order of $R_{\rm ISCO}$, all within an order of magnitude. Thus our results reinforce recent work presenting a near identical type of parametric comparison applied to BHB GX~339$-$4 presented by \cite{Sridhar2020}. The remarkable agreement between the expected size of the blackbody emitting disk (i.e. its inner radius) and the reflection constraint is encouraging in terms of the validity of the high-density reflection models for explaining the system properties. This connection, along with the consistency of our predicted ionizations with the measured value from reflection modeling, suggests we are arriving at an accurate physical representation of \u1630\ and its accretion flow properties.

We have also shown that in the soft state the reflection spectrum likely has a contribution from returning disk radiation. Our fits of the {\tt relxillNS} model to the \nustar/\swift-XRT data are superior to those made with {\tt reflionxHD}, which assumes the irradiating continuum is purely from a Comptonizing medium. A further consistency check provides us with supporting evidence for the presence of reflected returning disk radiation. The unabsorbed coronal IC fluxes derived from our Model c fits (Section~\ref{sec:nustar_fits}) are $3\times10^{-10}~{\rm erg~s^{-1}~cm^{-2}}$, $7\times10^{-10}~{\rm erg~s^{-1}~cm^{-2}}$ and $3\times10^{-11}~{\rm erg~s^{-1}~cm^{-2}}$ respectively. These fluxes are a mere 2\%, 3\% and 0.2 \% of the total source flux respectively. In addition, the fractional flux (of the total) of the reflection components {\tt relxillNS} are 4\%, 4\%, and 8\% respectively---we note that these estimates are derived using the best fit parameters, not the full confidence regions. \cite{Connors2020} showed that we would typically expect $\sim5\%$ of emitted disk photons to return to strike the disk for a black hole spin of $0.5$. Here we obtain relatively high BH spin ($a_{\star}=0.85\pm0.07$). In this case, the fraction of returning radiation can be several factors higher \citep{Connors2020,Wilkins2020}. 

Thus our spectral fitting results are consistent with predictions for maximally spinning BHs whereby the inner disk lies within a few $R_{\rm g}$ of the BH. There are strong caveats to this calculation though, since {\tt relxillNS} is not a physical model of returning radiation, and thus far we are relying on ray tracing of returning disk radiation which yields a {\it photon flux striking the disk} \citep{Wilkins2020}, not an {\it observed energy flux}, which is what we have derived from the fits. Therefore we stress that we only make this simple comparison to indicate that it may be reasonable to expect fluxes on the order of $\sim10\%$ in the reflected returning radiation component. Further tests are required to verify these numbers with full calculations of energy shifts in the returning disk emission, and this is work that is currently ongoing (Dauser et al., in preparation). 

It is worth comparing the results of reflection modeling of the \nustar\ observations with that of the \rxte\ observations presented in Section~\ref{sec:pca_fits}. Specifically, what impact would the inclusion of returning radiation (represented by the model {\tt relxillNS}) have on the parameter constraints we derived under the assumption of purely coronal disk illumination? Table~\ref{tab:nustar-params} shows that returning radiation can have an appreciable contribution to the total observed flux, and explain reflection features, when the source has a relatively minimal contribution from IC flux (with $f_{\rm sc}\le0.1$). Tables~\ref{tab:pca_fits_1998} and \ref{tab:pca_fits_2002} show we may expect returning radiation to contribute significantly once the source drops to hardness ratios below $\sim0.4$. Inclusion of the returning radiation component as the dominant reflector would likely result in lower values of the IC photon index, $\Gamma$, and potentially a necessary adjustment of the disk inclination. It is also possible we would necessarily have to adjust the BH spin, though this is difficult to predict without performing the fits.  

Finally, the constraints from reflection fitting imply a superabundance of iron in the disk, whereas we have assumed the disk wind has solar iron abundance ({\tt XSTAR}). Whilst this could skew our results, we note that the implied high abundances of iron in accretion disks via reflection modeling is still in the process of being understood (see, e.g., \citealt{Garcia2018b}). 

\subsection{Comparisons with Previous Work} \label{subsec:comparisons}

\cite{King2014} presented reflection modeling of the same 2013 \nustar\ observation of \u1630.  However, they adopted the model {\tt refbhb} \citep{Ross2007,Reis2008}, which accounts for the underlying blackbody emission of the disk propagating through the ionized upper layers of the disk. This has some similarities with reflected returning radiation, but there is a clear physical distinction. Thus it is difficult to make direct comparisons. However, our results are broadly consistent with those of \cite{King2014}, both in terms of the disk and coronal continuum parameters, and some key reflection parameters. For example we generally constrain the disk inclination to be in the range $60^{\circ}\mbox{--}70^{\circ}$ in high-density reflection fits, though we note that in our fits it was necessary to allow the emissivity index $q$ to vary freely. We also find similar disk density constraints ($\geq10^{21}~{\rm cm^{-3}}$). The difference in disk temperature (\citealt{King2014} find $kT_{\rm in }\sim1$~keV) likely occurs because \cite{King2014} adopt a reflection model which includes intrinsic underlying disk emission in the {\tt refbhb} model. Indeed \cite{King2014} find a similar disk temperature of $1.45$~keV when employing a simpler disk$+$power law model.

Our joint spectral fits to the joint \nustar/\swift-XRT observations during the soft state revealed that reflection from a disk blackbody performed better than reflection from a power law component. This echoes previous results found by \cite{DiazTrigo2014}. Direct comparison to their results is difficult, because they compare the {\tt relxill} model with a model which convolves any input continuum and performs the reflection calculation, {\tt rfxconv} \citep{Kolehmainen2011}. \cite{DiazTrigo2014} suggest, however, that the observed reflection features originate in the interaction of the disk emission with the ionized wind component, as opposed to reflection of photons off the disk (returning radiation). The rationale for this suggestion originates in the correlation between the strength of reflection features---such as Fe line equivalent width, and reflection fraction---and the column density of the warm absorber. This is a plausible explanation for the observed features, as has been found by recent works, e.g., \cite{Higginbottom2020}. Including these potential scattering and line emission effects in the warm absorber could affect our reflection modeling results, as such processes may mimic some of the reflection features, in particular the Fe line emission and Compton hump. Addressing these effects requires a much larger effort well outside the scope of this paper. 

\cite{Tomsick2005} explored \rxte\ observations of the $2002\mbox{--}2004$ outburst of \u1630, noting the extreme behaviors of \u1630 during transition to the soft state, and postulating that the source is close to the Eddington limit. They found that the inner disk temperature, $kT_{\rm in}$, varied between $2.7$ and $3.8$~keV during flaring in the intermediate state. The results presented in this work shed more light on the nature of the inner disk emission when reprocessing of coronal IC irradiation of a high-density disk is accounted for. As noted by \cite{Tomsick2005}, as the source exceeds $0.2L_{\rm Edd}$, electron scattering in the inner flow can heat the disk to such high temperatures. However, the effects of increased disk density are similar, with free-free heating increasing the temperature of the upper layers of the disk \citep{Garcia2016}. Thus these two processes are degenerate, which may partially explain the lower disk temperatures derived in our fits to the $2002\mbox{--}2004$ PCA observations---high-density reflection effects subsume some of the strong high-temperature disk emission discussed by \cite{Tomsick2005}. However, it should be noted that we are likely also deriving lower disk temperatures in the bright intermediate states due to the IC component we adopt ({\tt simplcut$\otimes$diskbb}), as opposed to the disconnected approach taken by \cite{Tomsick2005} ({\tt diskbb+powerlaw}). The former approach, which we took here, leads to lower disk temperatures due to the suppression of softer powerlaw flux when photon conservation is applied, as opposed to letting the power law fit independently of the disk (see \citealt{Steiner2009b} for a full discussion). 

Several recent articles have focused on high-density reflection modeling (e.g., \citealt{Tomsick2018,JJiang2019}). Unlike \cite{Tomsick2018}, who compared low density to high density reflection modeling of the BHB Cyg~X-1, we do not find strong evidence for a reduction in iron abundance constraints when adopting high density reflection models. This is in part due to the low statistical quality of the PCA data (we fixed $A_{\rm Fe}=5$ in all our fits), but applies to modeling of \nustar/\swift-XRT as well. Table~\ref{tab:nustar-params} shows that in our joint spectral fits, Model b.2 gives $A_{\rm Fe}>4.9$, close to the upper limit of the {\tt reflionxHD} model.

\cite{JJiang2019} fit high-density reflection models to \nustar\ and \swift\ observations of GX~$339-4$ during a relatively low-flux hard state and a bright soft state. They found disk densities in the lowest-flux hard state of $\sim10^{21}~{\rm cm^{-3}}$, and $\sim10^{19}~{\rm cm^{-3}}$ during a brighter soft state observation. We are not able to comment on the distinction between low-flux hard states and the soft state in our fits to \u1630, simply because of lack of good observational coverage of its rapid rise and transition. 

\subsection{Conclusions} \label{subsec:conclusions}
Two key conclusions arise from our results. Firstly, high-density modeling of the BHB \u1630 shows an improved consistency in the scalings of ionizing coronal IC flux and the corresponding disk ionization state given by the reflection model (in contrast to low-density reflection models). Secondly, echoing previous results found by \cite{Connors2020} and \cite{Wilkins2020}, during BHB soft states the irradiating continuum and resulting reflection component may be more like a reflected blackbody spectrum as opposed to reflection of the less dominant IC component.

We recommend exploring high-density reflection modeling of more sources, preferably in the bright hard state, with instruments such as \nustar\ and the {\it Neutron Star Interior Composition Explorer} ({\it NICER}), providing broadband X-ray coverage, sensitivity to the soft emission, and high spectral resolution. We need to build a large sample of such results in order to make broad comparisons with previous results using low-density reflection models. In addition, similar to the approach taken in this paper, one should check their reflection modeling results for consistency---the scalings of irradiating flux and the fit value of disk ionization should be consistent. 

Finally, in future work we will present more detailed models for returning radiation that are physically self-consistent (i.e., full relativistic ray tracing of returning photons from the disk, Dauser et al., in preparation). More self-consistent models will allow us to test the validity of returning radiation as a dominant reflection component, since the fraction and spectral shape of the returning photons depends strongly on BH spin and system geometry.

 \acknowledgements

We thank the referee for their comments, each of which facilitated the improvement of this manuscript. 

This work was partially supported under NASA contract No. NNG08FD60C and made use of data from the NuSTAR mission, a project led by the California Institute of Technology, managed by the Jet Propulsion Laboratory, and funded by the National Aeronautics and Space Administration. We thank the \nustar\ Operations, Software, and Calibration teams for support with the execution and analysis of these observations. This research has made use of the \nustar\ Data Analysis Software (NuSTARDAS), jointly developed by the ASI Science Data Center (ASDC, Italy) and the California Institute of Technology (USA).

 R.M.T.C. has been supported by NASA ADAP grant 80NSSC177K0515. J.A.G. acknowledges support from NASA APRA grant 80NSSC17K0345 and from the Alexander von Humboldt Foundation. VG is supported through the Margarete von Wrangell fellowship  by the ESF and the Ministry of Science, Research and the Arts  Baden-W\"urttemberg. JAT acknowledges partial support from NASA ADAP grant 80NSSC19K0586.
 
This research has made use of data, software and/or web tools obtained from the High Energy Astrophysics Science Archive Research Center (HEASARC), a service of the Astrophysics Science Division at NASA/GSFC and of the Smithsonian Astrophysical Observatory's High Energy Astrophysics Division. 

This research has made use of ISIS functions (ISISscripts) provided by 
ECAP/Remeis observatory and MIT (http://www.sternwarte.uni-erlangen.de/isis/).

\vspace{5mm}
\facilities{\rxte\ (PCA; \citealt{Jahoda1996}), \nustar\ \citep{Harrison:2013}, \swift-XRT \citep{Krimm2013}, HEASARC}

\software{{\tt XSPEC v.12.10.1s} \citep{Arnaud1996}, {\tt REFLIONX} \citep{Ross2005,Ross2007},
		 {\tt XILLVER} \citep{Garcia2010,Garcia2013}, {\tt RELXILL} (v1.3.3; \citealt{Garcia2014,Dauser2014}), {\tt PCACORR} \citep{Garcia2014}.}

\appendix
\section{PCA Spectral Fitting Results} \label{app:pcafits}
Tables~\ref{tab:pca_fits_1998} and~\ref{tab:pca_fits_2002} show the numerical parameter constraints from fits of Models a and b to PCA observations of \u1630 during $1998$ and $2002\mbox{--}2004$. The results of this analysis are discussed in detail in Section~\ref{sec:pca_fits}, and the values shown in Tables \ref{tab:pca_fits_1998} and \ref{tab:pca_fits_2002} correspond to the parameter trends shown in Figure~\ref{fig:par_trends}, but with some small distinctions due to rounding of errors. 


\def\OneLowDga{$1.88^{+0.04}_{-0.03}$}
\def\OneLowDfsc{$0.18^{+0.13}_{-0.06}$}
\def\OneLowDkTe{$30^{+70}_{-8}$}
\def\OneLowDTin{$0.31^{+0.28}_{-0.01}$}
\def\OneLowDndisk{$550^{+310}_{-50}$}
\def\OneLowDh{$<2.7$}
\def\OneLowDlogxi{$4.18^{+0.18}_{-0.08}$}
\def\OneLowDnrel{$9.9^{+0.9}_{-3.0}$}
\def\OneLowDchi{$54$}
\def\OneLowDnu{$68$}
\def\OneLowDchired{$0.79$}

\def\TwoLowDga{$2.095^{+0.002}_{-0.003}$}
\def\TwoLowDfsc{$<0.74$}
\def\TwoLowDkTe{$38^{+4}_{-3}$}
\def\TwoLowDTin{$0.92^{+0.02}_{-0.02}$}
\def\TwoLowDndisk{$6.6^{+0.1}_{-0.1}$}
\def\TwoLowDh{$5.5^{+5.1}_{-0.1}$}
\def\TwoLowDlogxi{$4.26^{+0.02}_{-0.03}$}
\def\TwoLowDnrel{$1.74^{+3.08}_{-0.04}$}
\def\TwoLowDchi{$56$}
\def\TwoLowDnu{$71$}
\def\TwoLowDchired{$0.79$}

\def\ThreeLowDga{$2.262^{+0.002}_{-0.003}$}
\def\ThreeLowDfsc{$<0.80$}
\def\ThreeLowDkTe{$70^{+10}_{-10}$}
\def\ThreeLowDTin{$0.92^{+0.02}_{-0.02}$}
\def\ThreeLowDndisk{$10.5^{+0.2}_{-5.0}$}
\def\ThreeLowDh{$5.0^{+0.1}_{-0.1}$}
\def\ThreeLowDlogxi{$4.23^{+0.03}_{-0.03}$}
\def\ThreeLowDDensity{$>60$}
\def\ThreeLowDnrel{$2.38^{+0.09}_{-0.09}$}
\def\ThreeLowDchi{$79$}
\def\ThreeLowDnu{$71$}
\def\ThreeLowDchired{$1.11$}

\def\FourLowDga{$2.3828^{+0.0015}_{-0.0007}$}
\def\FourLowDfsc{$0.806^{+0.006}_{-0.001}$}
\def\FourLowDkTe{$>210$}
\def\FourLowDTin{$0.766^{+0.002}_{-0.003}$}
\def\FourLowDndisk{$11.2^{+0.4}_{-0.4}$}
\def\FourLowDh{$3.75^{+0.08}_{-0.09}$}
\def\FourLowDlogxi{$>4.65$}
\def\FourLowDDensity{$>60$}
\def\FourLowDnrel{$6.22^{+0.09}_{-0.10}$}
\def\FourLowDchi{$86$}
\def\FourLowDnu{$71$}
\def\FourLowDchired{$1.22$}

\def\FiveLowDga{$2.584^{+0.004}_{-0.004}$}
\def\FiveLowDfsc{$0.905^{+0.007}_{-0.096}$}
\def\FiveLowDkTe{$>200$}
\def\FiveLowDTin{$0.916^{+0.005}_{-0.193}$}
\def\FiveLowDndisk{$12.8^{+0.7}_{-0.6}$}
\def\FiveLowDh{$5.5^{+0.3}_{-0.3}$}
\def\FiveLowDlogxi{$4.58^{+0.08}_{-0.08}$}
\def\FiveLowDDensity{$>60$}
\def\FiveLowDnrel{$4.7^{+2.4}_{-0.2}$}
\def\FiveLowDchi{$91$}
\def\FiveLowDnu{$68$}
\def\FiveLowDchired{$1.33$}

\def\SixLowDga{$2.802^{+0.004}_{-0.003}$}
\def\SixLowDfsc{$0.876^{+0.017}_{-0.005}$}
\def\SixLowDkTe{$>200$}
\def\SixLowDTin{$1.068^{+0.005}_{-0.054}$}
\def\SixLowDndisk{$11.4^{+0.1}_{-0.2}$}
\def\SixLowDh{$5.5^{+0.2}_{-0.2}$}
\def\SixLowDlogxi{$4.03^{+0.04}_{-0.07}$}
\def\SixLowDDensity{$>60$}
\def\SixLowDnrel{$5.6^{+34.7}_{-0.3}$}
\def\SixLowDchi{$81$}
\def\SixLowDnu{$68$}
\def\SixLowDchired{$1.19$}

\def\SevenLowDga{$2.822^{+0.001}_{-0.010}$}
\def\SevenLowDfsc{$0.6^{+0.1}_{-0.2}$}
\def\SevenLowDkTe{$>200$}
\def\SevenLowDTin{$1.059^{+0.017}_{-0.009}$}
\def\SevenLowDndisk{$5.58^{+0.07}_{-0.41}$}
\def\SevenLowDh{$5.48^{+1.02}_{-0.06}$}
\def\SevenLowDlogxi{$4.51^{+0.04}_{-0.03}$}
\def\SevenLowDDensity{$>60$}
\def\SevenLowDnrel{$7.41^{+2.17}_{-0.05}$}
\def\SevenLowDchi{$122$}
\def\SevenLowDnu{$69$}
\def\SevenLowDchired{$1.76$}

\def\EightLowDga{$2.39^{+0.08}_{-0.04}$}
\def\EightLowDfsc{$<0.4$}
\def\EightLowDkTe{$>100$}
\def\EightLowDTin{$1.01^{+0.01}_{-0.02}$}
\def\EightLowDndisk{$7^{+3}_{-1}$}
\def\EightLowDh{$6^{+5}_{-3}$}
\def\EightLowDlogxi{$4.25^{+0.09}_{-0.36}$}
\def\EightLowDDensity{$>60$}
\def\EightLowDnrel{$3^{+9}_{-2}$}
\def\EightLowDchi{$62$}
\def\EightLowDnu{$66$}
\def\EightLowDchired{$0.94$}

\def\NineLowDga{$2.27^{+0.05}_{-0.02}$}
\def\NineLowDfsc{$<0.1$}
\def\NineLowDkTe{$>100$}
\def\NineLowDTin{$1.044^{+0.005}_{-0.005}$}
\def\NineLowDndisk{$5.4^{+0.2}_{-0.1}$}
\def\NineLowDh{$6.8^{+2.5}_{-0.9}$}
\def\NineLowDlogxi{$4.13^{+0.02}_{-0.09}$}
\def\NineLowDDensity{$>60$}
\def\NineLowDnrel{$1.4^{+5.9}_{-0.6}$}
\def\NineLowDchi{$85$}
\def\NineLowDnu{$67$}
\def\NineLowDchired{$1.27$}


\def\OneHighDga{$1.78^{+0.05}_{-0.06}$}
\def\OneHighDfsc{$<0.9$}
\def\OneHighDkTe{$19^{+6}_{-3}$}
\def\OneHighDTin{$0.5^{+0.2}_{-0.2}$}
\def\OneHighDndisk{$30^{+90}_{-20}$}
\def\OneHighDh{$<100$}
\def\OneHighDlogxi{$4.47^{+0.09}_{-0.19}$}
\def\OneHighDDensity{$0.2^{+11.0}_{-0.1}$}
\def\OneHighDnrel{$2.0^{+2.3}_{-0.7}$}
\def\OneHighDchi{$51$}
\def\OneHighDnu{$67$}
\def\OneHighDchired{$0.76$}

\def\TwoHighDga{$1.79^{+0.12}_{-0.05}$}
\def\TwoHighDfsc{$<0.6$}
\def\TwoHighDkTe{$15^{+3}_{-1}$}
\def\TwoHighDTin{$1.01^{+0.03}_{-0.11}$}
\def\TwoHighDndisk{$3.5^{+3.6}_{-0.5}$}
\def\TwoHighDh{$<20$}
\def\TwoHighDlogxi{$3.99^{+0.36}_{-0.05}$}
\def\TwoHighDDensity{$>10$}
\def\TwoHighDnrel{$1.8^{+0.6}_{-0.8}$}
\def\TwoHighDchi{$52$}
\def\TwoHighDnu{$70$}
\def\TwoHighDchired{$0.75$}

\def\ThreeHighDga{$1.94^{+0.10}_{-0.09}$}
\def\ThreeHighDfsc{$<0.7$}
\def\ThreeHighDkTe{$18^{+12}_{-4}$}
\def\ThreeHighDTin{$0.98^{+0.05}_{-0.07}$}
\def\ThreeHighDndisk{$6^{+3}_{-1}$}
\def\ThreeHighDh{$15^{+13}_{-9}$}
\def\ThreeHighDlogxi{$4.0^{+0.4}_{-0.1}$}
\def\ThreeHighDDensity{$>2$}
\def\ThreeHighDnrel{$2.2^{+0.8}_{-0.6}$}
\def\ThreeHighDchi{$75$}
\def\ThreeHighDnu{$70$}
\def\ThreeHighDchired{$1.07$}

\def\FourHighDga{$1.88^{+0.04}_{-0.03}$}
\def\FourHighDfsc{$<0.6$}
\def\FourHighDkTe{$26^{+6}_{-4}$}
\def\FourHighDTin{$0.96^{+0.05}_{-0.05}$}
\def\FourHighDndisk{$7^{+4}_{-3}$}
\def\FourHighDh{$14^{+7}_{-5}$}
\def\FourHighDlogxi{$4.2^{+0.3}_{-0.2}$}
\def\FourHighDDensity{$21^{+24}_{-19}$}
\def\FourHighDnrel{$2.9^{+1.0}_{-0.7}$}
\def\FourHighDchi{$58$}
\def\FourHighDnu{$70$}
\def\FourHighDchired{$0.82$}

\def\FiveHighDga{$2.27^{+0.08}_{-0.12}$}
\def\FiveHighDfsc{$<0.7$}
\def\FiveHighDkTe{$22^{+26}_{-5}$}
\def\FiveHighDTin{$0.98^{+0.09}_{-0.07}$}
\def\FiveHighDndisk{$7^{+6}_{-2}$}
\def\FiveHighDh{$5^{+9}_{-3}$}
\def\FiveHighDlogxi{$4.4^{+0.2}_{-0.4}$}
\def\FiveHighDDensity{$13^{+66}_{-11}$}
\def\FiveHighDnrel{$6^{+13}_{-3}$}
\def\FiveHighDchi{$81$}
\def\FiveHighDnu{$67$}
\def\FiveHighDchired{$1.20$}

\def\SixHighDga{$2.32^{+0.23}_{-0.03}$}
\def\SixHighDfsc{$<0.4$}
\def\SixHighDkTe{$>30$}
\def\SixHighDTin{$1.14^{+0.02}_{-0.07}$}
\def\SixHighDndisk{$3.9^{+2.4}_{-0.2}$}
\def\SixHighDh{$6^{+3}_{-3}$}
\def\SixHighDlogxi{$3.98^{+0.23}_{-0.06}$}
\def\SixHighDDensity{$>10$}
\def\SixHighDnrel{$3.7^{+2.0}_{-0.8}$}
\def\SixHighDchi{$58$}
\def\SixHighDnu{$67$}
\def\SixHighDchired{$0.86$}

\def\SevenHighDga{$2.47^{+0.13}_{-0.03}$}
\def\SevenHighDfsc{$<0.4$}
\def\SevenHighDkTe{$>30$}
\def\SevenHighDTin{$1.01^{+0.02}_{-0.02}$}
\def\SevenHighDndisk{$5.2^{+2.5}_{-0.4}$}
\def\SevenHighDh{$5^{+5}_{-1}$}
\def\SevenHighDlogxi{$4.4^{+0.1}_{-0.1}$}
\def\SevenHighDDensity{$10^{+4}_{-7}$}
\def\SevenHighDnrel{$6.4^{+2.4}_{-0.9}$}
\def\SevenHighDchi{$48$}
\def\SevenHighDnu{$68$}
\def\SevenHighDchired{$0.70$}

\def\EightHighDga{$2.1^{+0.3}_{-0.1}$}
\def\EightHighDfsc{$<0.2$}
\def\EightHighDkTe{$>30$}
\def\EightHighDTin{$1.00^{+0.02}_{-0.04}$}
\def\EightHighDndisk{$8.2^{+2.6}_{-0.6}$}
\def\EightHighDh{$20^{+20}_{-10}$}
\def\EightHighDlogxi{$4.0^{+0.3}_{-0.2}$}
\def\EightHighDDensity{$>1$}
\def\EightHighDnrel{$2.0^{+1.6}_{-0.9}$}
\def\EightHighDchi{$55$}
\def\EightHighDnu{$65$}
\def\EightHighDchired{$0.85$}

\def\NineHighDga{$2.1^{+0.1}_{-0.5}$}
\def\NineHighDfsc{$<0.07$}
\def\NineHighDkTe{$>40$}
\def\NineHighDTin{$1.009^{+0.023}_{-0.008}$}
\def\NineHighDndisk{$6.4^{+0.2}_{-0.6}$}
\def\NineHighDh{$18^{+9}_{-8}$}
\def\NineHighDlogxi{$4.0^{+0.1}_{-0.2}$}
\def\NineHighDDensity{$10^{+12}_{-8}$}
\def\NineHighDnrel{$1.3^{+1.6}_{-0.3}$}
\def\NineHighDchi{$75$}
\def\NineHighDnu{$66$}
\def\NineHighDchired{$1.14$}

\begin{deluxetable*}{lccccccccc}
\tabletypesize{\scriptsize}
\tablecaption{Maximum likelihood estimates of all parameters in spectral fitting of \rxte-PCA observations of the $1998$ outburst of \u1630 with models (a) and (b). Model (a) enforces a fixed low-density reflection component ($n_{\rm e}=10^{15}~{\rm cm^{-3}}$), and model (b) has a variable disk density with $n_{\rm e} \le 10^{22}~{\rm cm^{-3}}$. \label{tab:pca_fits_1998}}
\tablecolumns{10}
\tablehead{
\colhead{Parameter} & 
\multicolumn{9}{c}{Hardness Ratio} \\
 &
\colhead{$0.78$} & 
\colhead{$0.67$} & 
\colhead{$0.63$} & 
\colhead{$0.58$} & 
\colhead{$0.54$} &
\colhead{$0.49$} &
\colhead{$0.46$} & 
\colhead{$0.41$} & 
\colhead{$0.32$}
}
\startdata
$R_{\rm in}~[R_{\rm ISCO}]$ &  \multicolumn{9}{c}{$1$}  \\
$a_{\rm \star}$ &  \multicolumn{9}{c}{$0.998$}  \\
$i~[^{\circ}]$ &  \multicolumn{9}{c}{$64$}  \\
$A_{\rm Fe}$ &  \multicolumn{9}{c}{$5$}  \\
$N_{\rm H}~[10^{22}~{\rm cm^{-2}}]$ &  \multicolumn{9}{c}{$14$}  \\
\hline
& \multicolumn{9}{c}{Model (a): {\tt TBabs(simplcut$\otimes$diskbb+relconvlp$\otimes$reflionx})} \\
\\
$\Gamma$ & \OneLowDga\ & \TwoLowDga\ & \ThreeLowDga\ & \FourLowDga\ & \FiveLowDga\ & \SixLowDga\ & \SevenLowDga\ & \EightLowDga\ & \NineLowDga\  \\
$f_{\rm sc}$ & \OneLowDfsc\ & \TwoLowDfsc & \ThreeLowDfsc\ & \FourLowDfsc\ & \FiveLowDfsc\ & \SixLowDfsc\ & \SevenLowDfsc\ & \EightLowDfsc\ & \NineLowDfsc\ \\
$T_{\rm in}$~[keV] & \OneLowDTin & \TwoLowDTin\ & \ThreeLowDTin\ & \FourLowDTin\ & \FiveLowDTin\ & \SixLowDTin\ & \SevenLowDTin\ & \EightLowDTin\ & \NineLowDTin\ \\
$kT_{\rm e}$~[keV] & \OneLowDkTe & \TwoLowDkTe\ & \ThreeLowDkTe\ & \FourLowDkTe\ & \FiveLowDkTe\ & \SixLowDkTe\ & \SevenLowDkTe\ & \EightLowDkTe\ & \NineLowDkTe\ \\
$N_{\rm disk}~[10^{2}]$ & \OneLowDndisk & \TwoLowDndisk\ & \ThreeLowDndisk\ & \FourLowDndisk\ & \FiveLowDndisk\ & \SixLowDndisk\  & \SevenLowDndisk\ & \EightLowDndisk\ & \NineLowDndisk\ \\
$h~[R_{\rm g}]$ & \OneLowDh\ & \TwoLowDh\ & \ThreeLowDh\ & \FourLowDh\ & \FiveLowDh\ & \SixLowDh\  & \SevenLowDh\ & \EightLowDh\ & \NineLowDh\ \\
 $\log{\xi}$~[${\rm erg~cm~s^{-1}}$] & \OneLowDlogxi\ & \TwoLowDlogxi\ & \ThreeLowDlogxi\ & \FourLowDlogxi\ & \FiveLowDlogxi\ & \SixLowDlogxi\ & \SevenLowDlogxi\ & \EightLowDlogxi\ & \NineLowDlogxi  \\
 $N_{\rm refl}~[10^{3}]$ & \OneLowDnrel & \TwoLowDnrel\ & \ThreeLowDnrel\ & \FourLowDnrel\ & \FiveLowDnrel\ & \SixLowDnrel\  & \SevenLowDnrel\ & \EightLowDnrel\ & \NineLowDnrel\ \\
 $n_{\rm e}~[{\rm cm^{-3}}]$ & \multicolumn{9}{c}{$10^{15}$} \\
 $\chi^2$ & \OneLowDchi\ & \TwoLowDchi\ & \ThreeLowDchi\ & \FourLowDchi\ & \FiveLowDchi\ & \SixLowDchi\   & \SevenLowDchi\ & \EightLowDchi\ & \NineLowDchi\ \\
 $\nu$ & \OneLowDnu\ & \TwoLowDnu\ & \ThreeLowDnu\ & \FourLowDnu\ & \FiveLowDnu\ & \SixLowDnu\  & \SevenLowDnu\ & \EightLowDnu\ & \NineLowDnu\ \\
 $\chi_{\nu}^2$ & \OneLowDchired & \TwoLowDchired & \ThreeLowDchired & \FourLowDchired & \FiveLowDchired & \SixLowDchired\  & \SevenLowDchired\ & \EightLowDchired\ & \NineLowDchired\ \\
 \hline
 & \multicolumn{9}{c}{Model (b): {\tt TBabs(simplcut$\otimes$diskbb+relconvlp$\otimes$reflionxHD})} \\
\\
$\Gamma$ & \OneHighDga\ & \TwoHighDga\ & \ThreeHighDga\ & \FourHighDga\ & \FiveHighDga\ & \SixHighDga\ & \SevenHighDga\ & \EightHighDga\ & \NineHighDga\  \\
$f_{\rm sc}$ & \OneHighDfsc\ & \TwoHighDfsc & \ThreeHighDfsc\ & \FourHighDfsc\ & \FiveHighDfsc\ & \SixHighDfsc\ & \SevenHighDfsc\ & \EightHighDfsc\ & \NineHighDfsc\ \\
$T_{\rm in}$~[keV] & \OneHighDTin & \TwoHighDTin\ & \ThreeHighDTin\ & \FourHighDTin\ & \FiveHighDTin\ & \SixHighDTin\ & \SevenHighDTin\ & \EightHighDTin\ & \NineHighDTin\ \\
$kT_{\rm e}$~[keV] & \OneHighDkTe & \TwoHighDkTe\ & \ThreeHighDkTe\ & \FourHighDkTe\ & \FiveHighDkTe\ & \SixHighDkTe\ & \SevenHighDkTe\ & \EightHighDkTe\ & \NineHighDkTe\ \\
$N_{\rm disk}~[10^{2}]$ & \OneHighDndisk & \TwoHighDndisk\ & \ThreeHighDndisk\ & \FourHighDndisk\ & \FiveHighDndisk\ & \SixHighDndisk\  & \SevenHighDndisk\ & \EightHighDndisk\ & \NineHighDndisk\ \\
$h~[R_{\rm g}]$ & \OneHighDh\ & \TwoHighDh\ & \ThreeHighDh\ & \FourHighDh\ & \FiveHighDh\ & \SixHighDh\  & \SevenHighDh\ & \EightHighDh\ & \NineHighDh\ \\
 $\log{\xi}$~[${\rm erg~cm~s^{-1}}$] & \OneHighDlogxi\ & \TwoHighDlogxi\ & \ThreeHighDlogxi\ & \FourHighDlogxi\ & \FiveHighDlogxi\ & \SixHighDlogxi\ & \SevenHighDlogxi\ & \EightHighDlogxi\ & \NineHighDlogxi  \\
 $N_{\rm refl}$ & \OneHighDnrel & \TwoHighDnrel\ & \ThreeHighDnrel\ & \FourHighDnrel\ & \FiveHighDnrel\ & \SixHighDnrel\  & \SevenHighDnrel\ & \EightHighDnrel\ & \NineHighDnrel\ \\
 $n_{\rm e}~[10^{20}~{\rm cm^{-3}}]$ & \OneHighDDensity\ &  \TwoHighDDensity\ & \ThreeHighDDensity\ & \FourHighDDensity\ & \FiveHighDDensity\ & \SixHighDDensity\ & \SevenHighDDensity\ & \EightHighDDensity\ & \NineHighDDensity\ \\
 $\chi^2$ & \OneHighDchi\ & \TwoHighDchi\ & \ThreeHighDchi\ & \FourHighDchi\ & \FiveHighDchi\ & \SixHighDchi\   & \SevenHighDchi\ & \EightHighDchi\ & \NineHighDchi\ \\
 $\nu$ & \OneHighDnu\ & \TwoHighDnu\ & \ThreeHighDnu\ & \FourHighDnu\ & \FiveHighDnu\ & \SixHighDnu\  & \SevenHighDnu\ & \EightHighDnu\ & \NineHighDnu\ \\
 $\chi_{\nu}^2$ & \OneHighDchired & \TwoHighDchired & \ThreeHighDchired & \FourHighDchired & \FiveHighDchired & \SixHighDchired\  & \SevenHighDchired\ & \EightHighDchired\ & \NineHighDchired\ \\
 \hline
\enddata
\tablecomments{The disk normalization is given by $N_{\rm disk} = (R_{\rm in}/D_{10})^2\cos\theta$, where $R_{\rm in}$ is the apparent inner disk in km, $D_{10}$ is the distance to the source in units of 10~kpc, and $\theta$ is the disk inclination. The total $\chi^2$ is shown for each fit, along with the degrees of freedom, $\nu$, and the reduced $\chi^2$, $\chi^2_{\nu}=\chi^2/\nu$. The ionization, $\log\xi$, is given by $L_{\rm irr}/n_{\rm e}R^2$, where $L_{\rm irr}$ is the ionizing luminosity, $n_{\rm e}$ is the gas density, and $R$ is the distance to the ionizing source. $N_{\rm refl}$ is the normalization of the reflection component, {\tt reflionxHD}. All other parameters are as described in the text.}
\end{deluxetable*}


\def\OneLowDga{$2.279^{+0.008}_{-0.006}$}
\def\OneLowDfsc{$0.4^{+0.6}_{-0.2}$}
\def\OneLowDkTe{$14.0^{+1.1}_{-0.4}$}
\def\OneLowDTin{$0.29^{+0.02}_{-0.01}$}
\def\OneLowDndisk{$6400^{+1277600}_{-200}$}
\def\OneLowDh{$40^{+10}_{-20}$}
\def\OneLowDlogxi{$>4.5$}
\def\OneLowDnrel{$0.05^{+0.38}_{-0.03}$}
\def\OneLowDchi{$78$}
\def\OneLowDnu{$62$}
\def\OneLowDchired{$1.25$}

\def\TwoLowDga{$2.301^{+0.004}_{-0.002}$}
\def\TwoLowDfsc{unconstrained}
\def\TwoLowDkTe{$11.9^{+0.3}_{-0.3}$}
\def\TwoLowDTin{$0.762^{+0.007}_{-0.022}$}
\def\TwoLowDndisk{$62^{+32}_{-7}$}
\def\TwoLowDh{$6^{+5}_{-2}$}
\def\TwoLowDlogxi{$3.3^{+0.4}_{-0.3}$}
\def\TwoLowDnrel{$3^{+12}_{-2}$}
\def\TwoLowDchi{$66$}
\def\TwoLowDnu{$63$}
\def\TwoLowDchired{$1.05$}

\def\ThreeLowDga{\nodata}
\def\ThreeLowDfsc{\nodata}
\def\ThreeLowDkTe{\nodata}
\def\ThreeLowDTin{\nodata}
\def\ThreeLowDndisk{\nodata}
\def\ThreeLowDh{\nodata}
\def\ThreeLowDlogxi{\nodata}
\def\ThreeLowDDensity{\nodata}
\def\ThreeLowDnrel{\nodata}
\def\ThreeLowDchi{\nodata}
\def\ThreeLowDnu{\nodata}
\def\ThreeLowDchired{\nodata}

\def\FourLowDga{\nodata}
\def\FourLowDfsc{\nodata}
\def\FourLowDkTe{\nodata}
\def\FourLowDTin{\nodata}
\def\FourLowDndisk{\nodata}
\def\FourLowDh{\nodata}
\def\FourLowDlogxi{\nodata}
\def\FourLowDDensity{\nodata}
\def\FourLowDnrel{\nodata}
\def\FourLowDchi{\nodata}
\def\FourLowDnu{\nodata}
\def\FourLowDchired{\nodata}

\def\FiveLowDga{$2.749^{+0.003}_{-0.003}$}
\def\FiveLowDfsc{$0.868^{+0.007}_{-0.096}$}
\def\FiveLowDkTe{$>220$}
\def\FiveLowDTin{$0.904^{+0.004}_{-0.093}$}
\def\FiveLowDndisk{$41.2^{+16.2}_{-0.4}$}
\def\FiveLowDh{$48^{+3}_{-3}$}
\def\FiveLowDlogxi{$4.03^{+0.11}_{-0.08}$}
\def\FiveLowDDensity{$>60$}
\def\FiveLowDnrel{$0.078^{+0.009}_{-0.007}$}
\def\FiveLowDchi{$81$}
\def\FiveLowDnu{$61$}
\def\FiveLowDchired{$1.33$}

\def\SixLowDga{$2.86^{+0.08}_{-0.12}$}
\def\SixLowDfsc{$0.51^{+0.05}_{-0.39}$}
\def\SixLowDkTe{$>100$}
\def\SixLowDTin{$1.222^{+0.017}_{-0.009}$}
\def\SixLowDndisk{$9^{+3}_{-3}$}
\def\SixLowDh{$6^{+3}_{-2}$}
\def\SixLowDlogxi{$4.1^{+0.6}_{-0.7}$}
\def\SixLowDDensity{$>60$}
\def\SixLowDnrel{$4.4^{+5.8}_{-0.1}$}
\def\SixLowDgabs{$0.017^{+0.008}_{-0.007}$}
\def\SixLowDchi{$98$}
\def\SixLowDnu{$60$}
\def\SixLowDchired{$1.64$}

\def\SevenLowDga{$2.74^{+0.06}_{-0.04}$}
\def\SevenLowDfsc{$0.24^{+0.03}_{-0.02}$}
\def\SevenLowDkTe{$>100$}
\def\SevenLowDTin{$1.424^{+0.004}_{-0.008}$}
\def\SevenLowDndisk{$4.27^{+0.11}_{-0.06}$}
\def\SevenLowDh{$<40$}
\def\SevenLowDlogxi{$<2.6$}
\def\SevenLowDDensity{$>60$}
\def\SevenLowDnrel{$<80$}
\def\SevenLowDgabs{$0.030^{+0.015}_{-0.009}$}
\def\SevenLowDchi{$83$}
\def\SevenLowDnu{$54$}
\def\SevenLowDchired{$1.54$}

\def\EightLowDga{$2.6^{+0.1}_{-0.6}$}
\def\EightLowDfsc{$0.10^{+0.04}_{-0.08}$}
\def\EightLowDkTe{$<300$}
\def\EightLowDTin{$1.362^{+0.016}_{-0.008}$}
\def\EightLowDndisk{$3.7^{+0.2}_{-0.2}$}
\def\EightLowDh{$<40$}
\def\EightLowDlogxi{$<4.6$}
\def\EightLowDDensity{$>60$}
\def\EightLowDnrel{$<0.4$}
\def\EightLowDgabs{$0.4^{+2.3}_{-0.3}$}
\def\EightLowDchi{$34$}
\def\EightLowDnu{$43$}
\def\EightLowDchired{$0.79$}

\def\NineLowDga{$>2.4$}
\def\NineLowDfsc{$<0.01$}
\def\NineLowDkTe{$<14$}
\def\NineLowDTin{$1.243^{+0.006}_{-0.016}$}
\def\NineLowDndisk{$3.6^{+0.2}_{-0.1}$}
\def\NineLowDh{$<50$}
\def\NineLowDlogxi{$2.0^{+0.3}_{-0.3}$}
\def\NineLowDDensity{$>60$}
\def\NineLowDnrel{$200^{+700}_{-100}$}
\def\NineLowDgabs{$0.3^{+0.5}_{-0.2}$}
\def\NineLowDchi{$33$}
\def\NineLowDnu{$33$}
\def\NineLowDchired{$1.01$}


\def\OneHighDga{$2.19^{+0.12}_{-0.02}$}
\def\OneHighDfsc{$0.19^{+0.28}_{-0.06}$}
\def\OneHighDkTe{$12.2^{+0.7}_{-1.2}$}
\def\OneHighDTin{$0.30^{+0.31}_{-0.01}$}
\def\OneHighDndisk{$8000^{+80000}_{-2000}$}
\def\OneHighDh{$<3$}
\def\OneHighDlogxi{$>4.5$}
\def\ThreeHighDDensity{$1.2^{+0.4}_{-1.1}$}
\def\OneHighDnrel{$30^{+6}_{-15}$}
\def\OneHighDchi{$73$}
\def\OneHighDnu{$61$}
\def\OneHighDchired{$1.20$}

\def\TwoHighDga{$2.22^{+0.01}_{-0.01}$}
\def\TwoHighDfsc{$0.18^{+0.14}_{-0.02}$}
\def\TwoHighDkTe{$12.1^{+0.8}_{-0.6}$}
\def\TwoHighDTin{$0.128^{+0.032}_{-0.004}$}
\def\TwoHighDndisk{$42000^{+10340000}_{-1000}$}
\def\TwoHighDh{$<20$}
\def\TwoHighDlogxi{$3.99^{+0.36}_{-0.05}$}
\def\ThreeHighDDensity{$>10$}
\def\TwoHighDnrel{$1.8^{+0.6}_{-0.8}$}
\def\TwoHighDchi{$46$}
\def\TwoHighDnu{$62$}
\def\TwoHighDchired{$0.74$}

\def\ThreeHighDga{$2.16^{+0.08}_{-0.02}$}
\def\ThreeHighDfsc{$<0.8$}
\def\ThreeHighDkTe{$13.2^{+0.4}_{-0.7}$}
\def\ThreeHighDTin{$1.05^{+0.01}_{-0.04}$}
\def\ThreeHighDndisk{$7.8^{+10.1}_{-0.3}$}
\def\ThreeHighDh{$26^{+18}_{-8}$}
\def\ThreeHighDlogxi{$4.44^{+0.11}_{-0.05}$}
\def\ThreeHighDDensity{$>60$}
\def\ThreeHighDnrel{$7.5^{+0.8}_{-1.6}$}
\def\ThreeHighDchi{$100$}
\def\ThreeHighDnu{$62$}
\def\ThreeHighDchired{$1.61$}

\def\FourHighDga{$2.53^{+0.06}_{-0.08}$}
\def\FourHighDfsc{$<0.9$}
\def\FourHighDkTe{$21^{+6}_{-4}$}
\def\FourHighDTin{$1.05^{+0.04}_{-0.05}$}
\def\FourHighDndisk{$12^{+68}_{-5}$}
\def\FourHighDh{$13^{+11}_{-4}$}
\def\FourHighDlogxi{$4.4^{+0.2}_{-0.2}$}
\def\FourHighDDensity{$>40$}
\def\FourHighDnrel{$8^{+13}_{-1}$}
\def\FourHighDchi{$64$}
\def\FourHighDnu{$62$}
\def\FourHighDchired{$1.03$}

\def\FiveHighDga{$2.615^{+0.003}_{-0.003}$}
\def\FiveHighDfsc{$0.78^{+0.05}_{-0.69}$}
\def\FiveHighDkTe{$34^{+2}_{-2}$}
\def\FiveHighDTin{$0.938^{+0.008}_{-0.008}$}
\def\FiveHighDndisk{$29.2^{+0.4}_{-0.4}$}
\def\FiveHighDh{$4.1^{+0.3}_{-0.3}$}
\def\FiveHighDlogxi{$4.10^{+0.07}_{-0.07}$}
\def\FiveHighDDensity{$0.72^{+0.08}_{-0.08}$}
\def\FiveHighDnrel{$8.4^{+63.1}_{-0.5}$}
\def\FiveHighDchi{$47$}
\def\FiveHighDnu{$60$}
\def\FiveHighDchired{$0.78$}

\def\SixHighDga{$2.47^{+0.03}_{-0.02}$}
\def\SixHighDfsc{$<0.05$}
\def\SixHighDkTe{$>100$}
\def\SixHighDTin{$1.20^{+0.02}_{-0.02}$}
\def\SixHighDndisk{$6.3^{+0.3}_{-0.4}$}
\def\SixHighDh{$7^{+9}_{-4}$}
\def\SixHighDlogxi{$4.47^{+0.06}_{-0.07}$}
\def\SixHighDDensity{$10^{+32}_{-4}$}
\def\SixHighDnrel{$4.7^{+1.3}_{-0.4}$}
\def\SixHighDgabs{$0.009^{+0.009}_{-0.007}$}
\def\SixHighDchi{$71$}
\def\SixHighDnu{$59$}
\def\SixHighDchired{$1.21$}

\def\SevenHighDga{$2.37^{+0.08}_{-0.03}$}
\def\SevenHighDfsc{$<0.04$}
\def\SevenHighDkTe{$>50$}
\def\SevenHighDTin{$1.403^{+0.008}_{-0.012}$}
\def\SevenHighDndisk{$3.60^{+0.11}_{-0.07}$}
\def\SevenHighDh{$>3$}
\def\SevenHighDlogxi{$>4.6$}
\def\SevenHighDDensity{$10^{+9}_{-4}$}
\def\SevenHighDnrel{$1.28^{+0.59}_{-0.09}$}
\def\SevenHighDgabs{$0.025^{+0.012}_{-0.009}$}
\def\SevenHighDchi{$72$}
\def\SevenHighDnu{$53$}
\def\SevenHighDchired{$1.35$}

\def\EightHighDga{$>2.2$}
\def\EightHighDfsc{$0.12^{+0.01}_{-0.05}$}
\def\EightHighDkTe{$<300$}
\def\EightHighDTin{$1.368^{+0.013}_{-0.009}$}
\def\EightHighDndisk{$3.7^{+0.1}_{-0.3}$}
\def\EightHighDh{$<60$}
\def\EightHighDlogxi{$<3.3$}
\def\EightHighDDensity{$<100$}
\def\EightHighDnrel{$<170$}
\def\EightHighDgabs{$0.2^{+0.6}_{-0.1}$}
\def\EightHighDchi{$33$}
\def\EightHighDnu{$42$}
\def\EightHighDchired{$0.79$}

\def\NineHighDga{$>2.7$}
\def\NineHighDfsc{$<0.01$}
\def\NineHighDkTe{$<140$}
\def\NineHighDTin{$1.243^{+0.007}_{-0.007}$}
\def\NineHighDndisk{$3.55^{+0.11}_{-0.09}$}
\def\NineHighDh{$<23$}
\def\NineHighDlogxi{$<1.5$}
\def\NineHighDDensity{$>10$}
\def\NineHighDnrel{$30^{+110}_{-20}$}
\def\NineHighDgabs{$0.16^{+0.24}_{-0.08}$}
\def\NineHighDchi{$29$}
\def\NineHighDnu{$32$}
\def\NineHighDchired{$0.92$}

\begin{deluxetable*}{lccccccccc}
\tabletypesize{\scriptsize}
\tablecaption{Maximum likelihood estimates of all parameters in spectral fitting of \rxte-PCA observations of the $2002\mbox{--}2004$ outburst of \u1630 with models (a) and (b). Model (a) enforces a fixed low-density reflection component ($n_{\rm e}=10^{15}~{\rm cm^{-3}}$), and model (b) has a variable disk density with $n_{\rm e} \le 10^{22}~{\rm cm^{-3}}$. \label{tab:pca_fits_2002}}
\tablecolumns{10}
\tablehead{
\colhead{Parameter} & 
\multicolumn{9}{c}{Hardness Ratio} \\
 &
\colhead{$0.63$} & 
\colhead{$0.62$} & 
\colhead{$0.60$} & 
\colhead{$0.53$} & 
\colhead{$0.48$} &
\colhead{$0.36$} &
\colhead{$0.27$} & 
\colhead{$0.21$} & 
\colhead{$0.13$}
}
\startdata
$R_{\rm in}~[R_{\rm ISCO}]$ &  \multicolumn{9}{c}{$1$}  \\
$a_{\rm \star}$ &  \multicolumn{9}{c}{$0.998$}  \\
$i~[^{\circ}]$ &  \multicolumn{9}{c}{$64$}  \\
$A_{\rm Fe}$ &  \multicolumn{9}{c}{$5$}  \\
$N_{\rm H}~[10^{22}~{\rm cm^{-2}}]$ &  \multicolumn{9}{c}{$14$}  \\
\hline
& \multicolumn{9}{c}{Model (a): {\tt TBabs(simplcut$\otimes$diskbb+relconvlp$\otimes$reflionx})} \\
\\
$\Gamma$ & \OneLowDga\ & \TwoLowDga\ & \ThreeLowDga\ & \FourLowDga\ & \FiveLowDga\ & \SixLowDga\ & \SevenLowDga\ & \EightLowDga\ & \NineLowDga\  \\
$f_{\rm sc}$ & \OneLowDfsc\ & \TwoLowDfsc & \ThreeLowDfsc\ & \FourLowDfsc\ & \FiveLowDfsc\ & \SixLowDfsc\ & \SevenLowDfsc\ & \EightLowDfsc\ & \NineLowDfsc\ \\
$T_{\rm in}$~[keV] & \OneLowDTin & \TwoLowDTin\ & \ThreeLowDTin\ & \FourLowDTin\ & \FiveLowDTin\ & \SixLowDTin\ & \SevenLowDTin\ & \EightLowDTin\ & \NineLowDTin\ \\
$kT_{\rm e}$~[keV] & \OneLowDkTe & \TwoLowDkTe\ & \ThreeLowDkTe\ & \FourLowDkTe\ & \FiveLowDkTe\ & \SixLowDkTe\ & \SevenLowDkTe\ & \EightLowDkTe\ & \NineLowDkTe\ \\
$N_{\rm disk}~[10^{2}]$ & \OneLowDndisk & \TwoLowDndisk\ & \ThreeLowDndisk\ & \FourLowDndisk\ & \FiveLowDndisk\ & \SixLowDndisk\  & \SevenLowDndisk\ & \EightLowDndisk\ & \NineLowDndisk\ \\
$h~[R_{\rm g}]$ & \OneLowDh\ & \TwoLowDh\ & \ThreeLowDh\ & \FourLowDh\ & \FiveLowDh\ & \SixLowDh\  & \SevenLowDh\ & \EightLowDh\ & \NineLowDh\ \\
 $\log{\xi}$~[${\rm erg~cm~s^{-1}}$] & \OneLowDlogxi\ & \TwoLowDlogxi\ & \ThreeLowDlogxi\ & \FourLowDlogxi\ & \FiveLowDlogxi\ & \SixLowDlogxi\ & \SevenLowDlogxi\ & \EightLowDlogxi\ & \NineLowDlogxi  \\
 $N_{\rm refl}~[10^{3}]$ & \OneLowDnrel & \TwoLowDnrel\ & \ThreeLowDnrel\ & \FourLowDnrel\ & \FiveLowDnrel\ & \SixLowDnrel\  & \SevenLowDnrel\ & \EightLowDnrel\ & \NineLowDnrel\ \\
 $n_{\rm e}~[{\rm cm^{-3}}]$ & \multicolumn{9}{c}{$10^{15}$} \\
  ${\rm Strength_{abs}}$ & \nodata\ & \nodata\ & \nodata\ & \nodata\ & \nodata\ & \SixLowDgabs\ & \SevenLowDgabs\ & \EightLowDgabs\ & \NineLowDgabs\ \\
 $\chi^2$ & \OneLowDchi\ & \TwoLowDchi\ & \ThreeLowDchi\ & \FourLowDchi\ & \FiveLowDchi\ & \SixLowDchi\   & \SevenLowDchi\ & \EightLowDchi\ & \NineLowDchi\ \\
 $\nu$ & \OneLowDnu\ & \TwoLowDnu\ & \ThreeLowDnu\ & \FourLowDnu\ & \FiveLowDnu\ & \SixLowDnu\  & \SevenLowDnu\ & \EightLowDnu\ & \NineLowDnu\ \\
 $\chi_{\nu}^2$ & \OneLowDchired & \TwoLowDchired & \ThreeLowDchired & \FourLowDchired & \FiveLowDchired & \SixLowDchired\  & \SevenLowDchired\ & \EightLowDchired\ & \NineLowDchired\ \\
 \hline
 & \multicolumn{9}{c}{Model (b): {\tt TBabs(simplcut$\otimes$diskbb+relconvlp$\otimes$reflionxHD})} \\
\\
$\Gamma$ & \OneHighDga\ & \TwoHighDga\ & \ThreeHighDga\ & \FourHighDga\ & \FiveHighDga\ & \SixHighDga\ & \SevenHighDga\ & \EightHighDga\ & \NineHighDga\  \\
$f_{\rm sc}$ & \OneHighDfsc\ & \TwoHighDfsc & \ThreeHighDfsc\ & \FourHighDfsc\ & \FiveHighDfsc\ & \SixHighDfsc\ & \SevenHighDfsc\ & \EightHighDfsc\ & \NineHighDfsc\ \\
$T_{\rm in}$~[keV] & \OneHighDTin & \TwoHighDTin\ & \ThreeHighDTin\ & \FourHighDTin\ & \FiveHighDTin\ & \SixHighDTin\ & \SevenHighDTin\ & \EightHighDTin\ & \NineHighDTin\ \\
$kT_{\rm e}$~[keV] & \OneHighDkTe & \TwoHighDkTe\ & \ThreeHighDkTe\ & \FourHighDkTe\ & \FiveHighDkTe\ & \SixHighDkTe\ & \SevenHighDkTe\ & \EightHighDkTe\ & \NineHighDkTe\ \\
$N_{\rm disk}~[10^{2}]$ & \OneHighDndisk & \TwoHighDndisk\ & \ThreeHighDndisk\ & \FourHighDndisk\ & \FiveHighDndisk\ & \SixHighDndisk\  & \SevenHighDndisk\ & \EightHighDndisk\ & \NineHighDndisk\ \\
$h~[R_{\rm g}]$ & \OneHighDh\ & \TwoHighDh\ & \ThreeHighDh\ & \FourHighDh\ & \FiveHighDh\ & \SixHighDh\  & \SevenHighDh\ & \EightHighDh\ & \NineHighDh\ \\
 $\log{\xi}$~[${\rm erg~cm~s^{-1}}$] & \OneHighDlogxi\ & \TwoHighDlogxi\ & \ThreeHighDlogxi\ & \FourHighDlogxi\ & \FiveHighDlogxi\ & \SixHighDlogxi\ & \SevenHighDlogxi\ & \EightHighDlogxi\ & \NineHighDlogxi  \\
 $N_{\rm refl}$ & \OneHighDnrel & \TwoHighDnrel\ & \ThreeHighDnrel\ & \FourHighDnrel\ & \FiveHighDnrel\ & \SixHighDnrel\  & \SevenHighDnrel\ & \EightHighDnrel\ & \NineHighDnrel\ \\
 $n_{\rm e}~[10^{20}~{\rm cm^{-3}}]$ & \OneHighDDensity\ &  \TwoHighDDensity\ & \ThreeHighDDensity\ & \FourHighDDensity\ & \FiveHighDDensity\ & \SixHighDDensity\ & \SevenHighDDensity\ & \EightHighDDensity\ & \NineHighDDensity\ \\
 ${\rm Strength_{abs}}$ & \nodata\ & \nodata\ & \nodata\ & \nodata\ & \nodata\ & \SixHighDgabs\ & \SevenHighDgabs\ & \EightHighDgabs\ & \NineHighDgabs\ \\
 $\chi^2$ & \OneHighDchi\ & \TwoHighDchi\ & \ThreeHighDchi\ & \FourHighDchi\ & \FiveHighDchi\ & \SixHighDchi\   & \SevenHighDchi\ & \EightHighDchi\ & \NineHighDchi\ \\
 $\nu$ & \OneHighDnu\ & \TwoHighDnu\ & \ThreeHighDnu\ & \FourHighDnu\ & \FiveHighDnu\ & \SixHighDnu\  & \SevenHighDnu\ & \EightHighDnu\ & \NineHighDnu\ \\
 $\chi_{\nu}^2$ & \OneHighDchired & \TwoHighDchired & \ThreeHighDchired & \FourHighDchired & \FiveHighDchired & \SixHighDchired\  & \SevenHighDchired\ & \EightHighDchired\ & \NineHighDchired\ \\
 \hline
\enddata
\tablecomments{The disk normalization is given by $N_{\rm disk} = (R_{\rm in}/D_{10})^2\cos\theta$, where $R_{\rm in}$ is the apparent inner disk in km, $D_{10}$ is the distance to the source in units of 10~kpc, and $\theta$ is the disk inclination. The total $\chi^2$ is shown for each fit, along with the degrees of freedom, $\nu$, and the reduced $\chi^2$, $\chi^2_{\nu}=\chi^2/\nu$. The ionization, $\log\xi$, is given by $L_{\rm irr}/n_{\rm e}R^2$, where $L_{\rm irr}$ is the ionizing luminosity, $n_{\rm e}$ is the gas density, and $R$ is the distance to the ionizing source. $N_{\rm refl}$ is the normalization of the reflection component, {\tt reflionxHD}. ${\rm Strength_{abs}}$ is the strength of the Gaussian absorption line, with centroid energy $E_{\rm abs}=6.9$~keV, and width $\sigma_{\rm abs}=10$~eV, both frozen parameters. All other parameters are as described in the text. Observations 80117-01-03-00G ($\rm {HR}=0.60$) and 80117-01-07-01 (${\rm HR}=0.53$) are not included in Model (a) spectral fitting results due to poor fits, and thus unreliable parameter constraints.}
\end{deluxetable*}

\bibliographystyle{aasjournal}
\bibliography{references}
%
%
%
%

\end{document}